\documentclass[preprint,prd,aps,amsmath,nofootinbib,superscriptaddress,tightenlines]{revtex4}

\usepackage[hypertex]{hyperref}
\usepackage{graphicx}
\usepackage{bm}
\usepackage{xspace}
\usepackage{slashed}

  \newcount\hour \newcount\minute
  \hour=\time \divide \hour by 60
  \minute=\time
  \newcommand{\mydate}{\ \today \ - \number\hour :\ifnum \minute<10 0\fi 
\number\minute}

\def\Dslash{D\!\!\!\!\slash}

\def\vslash{v\!\!\!\slash}
\def\OMIT#1{}

\newcommand{\nslash}{\slashed{n}}
\newcommand{\bnslash}{\slashed{\bar n}}

\newcommand{\nn}{\nonumber} 

\newcommand{\bn}{{\bar n}}
\newcommand{\bea}{\begin{eqnarray}}
\newcommand{\eea}{\end{eqnarray}}

\newcommand{\bnP}{\bar {\cal P}}

\newcommand{\cP}{{\cal P}}

\newcommand{\mcdot}{\!\cdot\!}

\newcommand{\SCETa}{\ensuremath{{\rm SCET}_{\rm I}}\xspace}
\newcommand{\SCETb}{\ensuremath{{\rm SCET}_{\rm II}}\xspace}
\newcommand{\muplus}{\mu_+}
\newcommand{\muminus}{\mu_-}

\begin{document}
\setlength\baselineskip{17pt}


\preprint{ \vbox{ \hbox{MIT-CTP 3751} \hbox{LBNL-60196} \hbox{CMU-HEP-06-08}  
 \hbox{hep-ph/0607001} } 
}

\title{\boldmath 
Power Corrections in Charmless Nonleptonic $B$ Decays:\\[3pt]
Annihilation is Factorizable and Real}

\vspace*{1cm}

\author{Christian M.\ Arnesen} 
\affiliation{Center for Theoretical Physics, Laboratory
for Nuclear Science, Massachusetts \\
Institute of Technology, Cambridge, MA 02139}

\author{Zoltan Ligeti}
\affiliation{Ernest Orlando Lawrence Berkeley National Laboratory,
University of California, \\ 
Berkeley, CA 94720}
\affiliation{Center for Theoretical Physics, Laboratory
for Nuclear Science, Massachusetts \\
Institute of Technology, Cambridge, MA 02139}

\author{Ira Z.\ Rothstein}
\affiliation{Department of Physics, Carnegie Mellon University,
  Pittsburgh, PA 15213\vspace*{0.5cm}}

\author{Iain W.\ Stewart\vspace{0.4cm}}
\affiliation{Center for Theoretical Physics, Laboratory
for Nuclear Science, Massachusetts \\
Institute of Technology, Cambridge, MA 02139}

\begin{abstract}\vspace*{0.2cm}

  We classify $\Lambda_{\rm QCD}/m_b$ power corrections to nonleptonic $B\to M_1
  M_2$ decays, where $M_{1,2}$ are charmless non-isosinglet mesons.  Using
  recent developments in soft-collinear effective theory, we prove that the
  leading contributions to annihilation amplitudes of order $\alpha_s(m_b)
  \Lambda_{\rm QCD}/m_b$ are real.  The leading annihilation amplitudes depend
  on twist-2 and the twist-3 three parton distributions.  A complex
  nonperturbative parameter from annihilation first appears at ${\cal
    O}\big[\alpha_s^2(\sqrt{\Lambda m_b})\Lambda_{\rm QCD}/m_b \big]$.
  ``Chirally enhanced'' contributions are also factorizable and real at lowest
  order. Thus, incalculable strong phases are suppressed in annihilation
  amplitudes, unless the $\alpha_s(\sqrt{\Lambda m_b})$ expansion breaks down.
  Modeling the distribution functions, we find that $(11\pm 9)\%$ and $(15\pm
  11)\%$ of the absolute values of the measured $\bar B^0\to K^-\pi^+$ and
  $B^-\to K^-K^0$ penguin amplitudes come from annihilation. This is consistent
  with the expected size of power corrections.

\end{abstract}

\maketitle

\section{Introduction}

Nonleptonic charmless $B$ decays are important probes of the standard model.
They are sensitive to the $CP$ violating phase $\gamma$ (or $\alpha$) via the
interference of tree and penguin contributions, and to possible new physics that
could modify the penguin amplitudes.  They also provide a powerful laboratory to
study strong interactions, the understanding of which is crucial if one is to
claim sensitivity to new physics in these decays.

The theory of nonleptonic $B$ decays underwent important progress in the last
few years.  Factorization theorems for $B\to M M'$ decays have been proven to
all orders in $\alpha_s$ at leading order in $\Lambda/m_b$, for decays when $M$
is a light (charmless) meson and $M'$ is either charmed or
charmless~\cite{Bauer:2001cu,BBNS,Mantry:2003uz,Bauer:2004tj,chay}. Here
$\Lambda\sim \Lambda_{\rm QCD}\sim 500\,{\rm MeV}$ denotes a typical hadronic
scale. An important difference between the various approaches to making
predictions for the charmless $B\to M_1 M_2$ decay
rates~\cite{Keum,Keum:2000wi,Lu:2000em,BBNS,BBNS2,BN,Bauer:2004tj,Bauer:2005kd,Colangelo,Ciuchini}
is how certain ${\cal O}(\Lambda/m_b)$ power suppressed corrections are treated.
In particular, it was observed that so-called annihilation diagrams (as in
Fig.~\ref{fig:Ann}) give rise to divergent convolution integrals if one attempts
calculating them using conventional factorization techniques~\cite{Keum}.  In
the KLS (or pQCD) approach~\cite{Keum}, these are rendered finite by $k_\perp$
dependences, which effectively cut off the endpoints of the meson distribution
functions.  KLS found large imaginary parts from the jet scale,
$\sqrt{m_b\Lambda}$, from propagators via ${\rm Im}\,[x m_b^2 -
k_\perp^2+i\epsilon]^{-1}= -\pi\delta(xm_b^2-k_\perp^2)$~\cite{Li:conf}. They
also found that for the physical value of $m_b$ the power suppression of these
terms relative to the leading contributions was not very significant.  In the
BBNS (or QCDF) approach~\cite{BBNS2,BN,BBNS}, the divergent convolutions are
interpreted as signs of infrared sensitive contributions, and are modeled by
complex parameters, $X_A=\int_0^1 dy/y = (1 + \rho_A e^{i\varphi_A})
\ln(m_B/\Lambda)$, with $\rho_A\le 1$ and an unrestricted strong phase
$\varphi_A$. In Ref.~\cite{Feldmann:2004mg} annihilation diagrams were
investigated in the soft-collinear effective theory (SCET)~\cite{SCET} and
parameterized by a complex amplitude. When annihilation is considered in $SU(3)$
flavor analyses a complex parameter is also used~\cite{su3}. In the absence of a
factorization theorem for annihilation contributions, a dimensional analysis
based parameterization with $\Lambda/m_b$ magnitude and unrestricted strong
phases is a reasonable way of estimating the uncertainty. In order not to
introduce model dependent correlations, a new parameter could be used for each
independent channel.

It was recently shown by Manohar and Stewart~\cite{Manohar:2006nz} that properly
separating the physics at different momentum scales removes the divergences,
giving well defined results for convolution integrals through a new type of
factorization which separates modes in their invariant mass and rapidity. The
analysis involves a minimal subtraction with the zero-bin method to avoid double
counting rapidity regions, and with the regulation and subtraction of
divergences for large $p^+$ and $p^-$ momenta that behave like ultraviolet
divergences. Additional subtractions would correspond to scheme dependent terms,
so the minimal subtraction is the usual and simplest choice.  We refer to this
as MS factorization.\footnote{Over the objection of one of the authors.} In this
paper we classify annihilation contributions to $B\to M_1 M_2$ decays and
demonstrate how this rapidity factorization works for the leading terms of order
${\cal O}[\alpha_s(m_b)\Lambda/m_b]$. These leading order annihilation
contributions are real despite the presence of endpoint divergences.  We also
classify which terms can involve a nonperturbative complex hadronic parameter,
and show that they first show up for annihilation at higher order in
perturbation theory, ${\cal O}[\alpha_s^2(\sqrt{m_b \Lambda})\, \Lambda/m_b]$.

Our analysis demonstrates that while certain annihilation contributions are only
sensitive to the hard short distance scale $\mu^2\sim m_b^2$ (local
annihilation), there exist other annihilation contributions that start at the
same order in $\alpha_s$ and $1/m_b$ and are sensitive to the intermediate scale
$\mu^2\sim m_b\Lambda$ (hard-collinear annihilation terms). The leading local
annihilation terms involve $f_B$ and a modified type of twist-2 distribution
functions, while the leading hard-collinear terms have twist-3 meson
distributions. In this work we perform matching calculations for the two-body
distributions that require rapidity factorization.  The calculation of the
leading amplitude involving the three body functions is given in a separate
publication~\cite{Arnesen2}, however we review the numerical results here.

An interesting set of power corrections are those proportional to $\mu_P$ where
$\mu_\pi = m_\pi^2/(m_u + m_d)$ and $\mu_K = m_K^2/(m_u +
m_s)$~\cite{Shifman:1975tn}.  For kaons and pions $\mu_P \sim 2\,$GeV, so
corrections proportional to $\mu_P/m_b$ can be sizable, and were labeled
``chirally enhanced" in Ref.~\cite{BBNS,BBNS2}.  In the chiral limit $\mu_P
\propto \Lambda_{\chi}$, where $\Lambda_{\chi}$ is the chiral symmetry breaking
scale, so the enhancement is not parametric, and comes from the fact that
$\Lambda_{\chi} > \Lambda_{\rm QCD}$.  In the BBNS approach these
$\Lambda^2/m_b^2$ annihilation power corrections are included along with the
leading order terms, and when they multiply divergent convolutions they are
described by complex parameters. Below we show that, much like the lowest order
annihilation contributions, these terms are also real and factorizable.

In section~\ref{sec:classify} we review the leading order factorization theorem,
and classify power corrections to $B\to M_1 M_2$, with a focus on annihilation
amplitudes.  In section~\ref{sec:local} a factorization theorem is derived for
local annihilation amplitudes at order $\Lambda/m_b$ for final states not
involving isosinglets (given in Eq.~(\ref{Aann})). These amplitudes start at
${\cal O}(\alpha_s(m_b))$ and involve $f_B$ and a modified type of twist-2 meson
distributions.  The extension to chirally enhanced local annihilation terms is
considered in section~\ref{sec:chiral}. In section~\ref{sec:phase} we study
annihilation amplitudes from time-ordered products, and classify complex
contributions generated at the hard scale $m_b$, the intermediate scale
$\sqrt{m_b\Lambda}$, and the nonperturbative scale $\Lambda$. Our results give
absolute predictions for the annihilation amplitudes in $B\to PP,\, PV,\, VV$
channels, given the meson distribution functions as inputs, which are studied in
Section~\ref{sec:phen}.  This section also discusses the implications of our
results for models of annihilation used in the literature, and a numerical
analysis of the annihilation amplitudes in $\bar B\to K\pi$ and $\bar B\to K\bar
K$.  Appendix~\ref{AppA} gives the derivation of a two-dimensional convolution
formula with overlapping zero-bin subtractions.


\section{Annihilation Contributions in SCET}
\label{sec:classify}


We use $M$ to denote a charmless pseudoscalar or vector meson ($\pi$, $K$,
$\rho$, $\ldots$). The relevant scales in $B\to M_1 M_2$ decays are $m_W$,
$m_b$, $E \approx m_B/2$, $m_c$, the jet scale $\sqrt{E\Lambda}$, and the
nonperturbative scale $\Lambda$.  Here $E$ is the energy of the light mesons,
which is much greater than their masses, $m_{M_{1,2}}\sim \Lambda$. To simplify
notation, we denote by $m_b$ hereafter the expansion in all hard scales,
$\{m_b,E,m_c\}$.  The decays $B \to M_1 M_2$ are mediated by the weak $\Delta
B=1$ effective Hamiltonian, which has $\Delta S=0$ terms for $\bar b \to \bar d
q_1 \bar q_2$ transitions and $\Delta S=1$ terms for $\bar b \to \bar s q_1 \bar
q_2$. For $\Delta S=0$ it reads
\begin{equation} \label{Hw}
H_W = \frac{G_F}{\sqrt{2}} \sum_{p=u,c} V_{pb} V^*_{pd}\,
  \Big( C_1 O_1^p + C_2 O_2^p 
  + \!\sum_{i=3}^{10,7\gamma,8g}\! C_i O_i \Big),
\end{equation}
where the operators are
\begin{align}\label{fullops}
 O_1^u  &= (\bar u b)_{V\!-\!A}\,
  (\bar d u)_{V\!-\!A}, 
 & O_2^u &= (\bar u_{\beta} b_{\alpha})_{V\!-\!A}\,
  (\bar d_{\alpha} u_{\beta})_{V\!-\!A} \,, \nn \\
 O_1^c  &= (\bar c b)_{V\!-\!A}\,
  (\bar d c)_{V\!-\!A}, 
 & O_2^c &= (\bar c_{\beta} b_{\alpha})_{V\!-\!A}\,
  (\bar d_{\alpha} c_{\beta})_{V\!-\!A} \,, \nn\\
 O_{3}  &=  {\textstyle \sum}_{q'} (\bar d b)_{V\!-\!A}\,
  (\bar q' q')_{V\! - \!A} \,,
 & O_{4} & =  {\textstyle \sum}_{q'} (\bar d_{\beta} b_{\alpha})_{V\!-\!A}\,
  (\bar q'_{\alpha} q'_{\beta})_{V\! - \!A} \,, \nn \\
 O_{5}  &= {\textstyle \sum}_{q'} (\bar d b)_{V\!-\!A}\,
  (\bar q' q')_{V\! + \!A} \,, 
 & O_6 &= {\textstyle \sum}_{q'} (\bar d_{\beta} b_{\alpha})_{V\!-\!A}\,
  (\bar q'_{\alpha} q'_{\beta})_{V\! + \!A} \,, \nn \\
 O_{7}  &= {\textstyle \sum}_{q'} \frac{3e_{q'}}{2}\, (\bar d b)_{V\!-\!A}\,
  (\bar q' q')_{V\! + \!A}\,,
 & O_8 & = {\textstyle \sum}_{q'} \frac{3e_{q'}}{2}\,
  (\bar d_{\beta} b_{\alpha})_{V\!-\!A}\,
  (\bar q'_{\alpha} q'_{\beta})_{V\! + \!A} \,, \nn \\
 O_{9}  &= {\textstyle \sum}_{q'} \frac{3e_{q'}}{2}\,
  (\bar d b)_{V\!-\!A}\, (\bar q' q')_{V\! - \!A}\,, 
 & O_{10} & = {\textstyle \sum}_{q'} \frac{3e_{q'}}{2}\,
  (\bar d_{\beta} b_{\alpha})_{V\!-\!A}\,
  (\bar q'_{\alpha} q'_{\beta})_{V\!-\!A} \,, \nn \\
 O_{7\gamma}  &=  -\frac{e}{8\pi^2}\, m_b\, \bar d\, \sigma^{\mu\nu}
   F_{\mu\nu} (1\!+\! \gamma_5)  b \,,
 & O_{8g}  &=  -\frac{g}{8\pi^2}\, m_b\, \bar d\, \sigma^{\mu\nu}
  G_{\mu\nu}^a T^a (1\!+\!\gamma_5) b \,.
\end{align}
Here $O_{1,2}^u$ and $O_{1,2}^c$ are current-current operators,
$\alpha$ and $\beta$ are color indices, $O_{3-6}$ are penguin
operators and $O_{7-10}$ are electroweak penguin operators, with a sum
over $q'=u,d,s,c,b$ flavors, and electric charges $e_{q'}$.  Results for $\Delta S=1$ transitions are
obtained by replacing $d\to s$ in Eqs.~(\ref{Hw}) and (\ref{fullops}),
and likewise in the equations below.  The coefficients in
Eq.~(\ref{Hw}) are known at NLL order~\cite{fullWilson} (we have
$O^p_{1}\leftrightarrow O^p_{2}$ relative to~\cite{fullWilson}). In
the NDR scheme, taking $\alpha_s(m_Z)=0.118$ and $m_b=4.8\,{\rm GeV}$,
\begin{eqnarray} \label{Ci}
 && 
C_{1-10}(m_b) = \big\{
  1.080\,,\
  -0.177\,,\
  0.011\,,\
 -0.033\,,\
  0.010\,,\
 -0.040 \,, 
   \nn\\
 && \hspace{2.5cm}
  4.9 \!\times\! 10^{-4} \,,\
  4.6 \!\times\! 10^{-4} \,,\
  -9.8 \!\times\! 10^{-3} \,,\
  1.9 \!\times\! 10^{-3} \big\} \,.
\end{eqnarray}

To define what we mean by annihilation amplitudes we use the contraction
amplitudes $A_1$, $A_2$, $P_3$, $P_3^{\rm GIM}$ in the full electroweak theory
from Ref.~\cite{Buras:1998ra} (which thus includes penguin annihilation). These
amplitudes are scheme and scale independent and correspond to Feynman diagrams
with a Wick contraction between the spectator flavor in the initial state and a
quark in the operators $O_i$. Using SCET these annihilation amplitudes can be
proven to be suppressed by $\Lambda/m_b$ to all orders in
$\alpha_s$~\cite{Bauer:2004tj}. These contributions differ from
emission-annihilation amplitudes, $E\!A_{1}$ and $E\!A_{2}$, which involve at
least one isosinglet meson. As demonstrated in
Refs.~\cite{BN,Williamson:2006hb}, $E\!A_{1,2}$ occur at leading order in the
power expansion. We focus on isodoublet and isotriplet final states,
so ignore the $E\!A_{1,2}$ amplitudes hereafter.

To separate the mass scales occurring below $m_b$ we need to match $H_W$ onto
operators in SCET.  The nonperturbative degrees of freedom are soft quarks and
gluons for the $B$-meson, $n$-collinear quarks and gluons for one light meson,
and $\bn$-collinear fields for the other light meson, as defined
in~\cite{bfprs}.  Expanding in $\Lambda/m_b$ gives
\begin{align} \label{Aexpn}
\langle M_1 M_2 | H_W | B \rangle
  & =  A^{(0)} + A_{c\bar c} + A^{(1)}_{ann}
  + A^{(1)}_{rest} + \ldots \nn\\
& = \frac{G_F m_B f_{M_1} f_{M_2} f_B}
  {\sqrt{2}\,  \Lambda_0 } 
  \Big[ \hat A^{(0)} + \hat A_{c\bar c} + \hat A^{(1)}_{ann}  
  + \hat A^{(1)}_{rest} + \ldots \Big] \,.
\end{align}
In the second line we switched to dimensionless amplitudes $\hat A$ by pulling
out a prefactor with the correct $\Lambda^{5/2} m_b^{1/2}$ scaling. Here
$\Lambda_0=500\,{\rm MeV}$ represents a $B$-meson scale that is ${\cal
  O}(\Lambda_{\rm QCD})$.  Taking $\eta=\Lambda/m_b$ we have the leading order
amplitude $\hat A^{(0)}= {\cal O}(\eta^0)$, and the subleading amplitude $\hat
A^{(1)}= \hat A_{ann}^{(1)}+\hat A_{rest}^{(1)}= {\cal O}(\eta^1)$, which we
have split into the annihilation amplitude $\hat A_{ann}^{(1)}$ and the
remainder $\hat A_{rest}^{(1)}$. The amplitude $\hat A_{c\bar c}$ in
Eq.~(\ref{Aexpn}), denotes contributions from long-distance charm effects in all
amplitudes, while perturbative charm loops contribute in the amplitudes
$A^{(0)}$ and $A^{(1)}$.\footnote{$\hat A_{c\bar c}$ has the $c$-fields in
  $O_{1,2}^c$ and $O_{3-10}$ replaced by nonrelativistic
  fields~\cite{Bauer:2004tj}, and is suppressed by at least their relative
  velocity, $v\sim 0.3-0.5$. The possibility of large nonperturbative charm loop
  contributions was first discussed in Refs.~\cite{Colangelo,Ciuchini}, and the
  size of these terms remains controversial~\cite{Beneke:2004bn,Bauer:2005wb}.}

There are two formally large scales, $m_b \gg \sqrt{m_b\Lambda} \gg \Lambda$,
which we will refer to as the hard scale $\mu_h \sim m_b$, and intermediate or
hard-collinear scale $\mu_i \sim \sqrt{m_b\Lambda}$.  These scales can be
integrated out one-by-one~\cite{Bauer:2002aj} with effective theories \SCETa and
\SCETb.  Integrating out $m_b$ requires matching the $O_i$ onto a series of
operators in \SCETa, $Q^{(j)}\sim \lambda^j$ where the \SCETa power counting
parameter $\lambda=\eta^{1/2}=\sqrt{\Lambda/m_b}$.  To obtain contributions to
$B\to M_1M_2$, we require an odd number of ultrasoft (usoft) light quarks
$q_{us}$, two or more $n$-collinear fields, and two or more $\bn$-collinear
fields, where $n^2=\bn^2=0$.

We briefly review results from Refs.~\cite{chay,Bauer:2004tj} for the leading
amplitude $A^{(0)}$ for $B\to M_1 M_2$. Here we have weak operators
$Q_{1d-6d}^{(0)}\sim \lambda^6$, $Q_{1d-8d}^{(1)}\sim \lambda^{7}$ with no
$q_{us}$'s, taken in time-ordered products with an usoft-collinear quark
Lagrangian, ${\cal L}_{\xi q}^{(j)}\sim \lambda^{j}$ for $j=1,2$, which has one
$q_{us}$.  We denote other subleading Lagrangians by ${\cal L}^{(j)}$, and list
the ${\cal O}(\lambda^{7})$ and ${\cal O}(\lambda^{8})$ time-ordered products
for $A^{(0)}$ in Table~\ref{table0}.  Matching these time-ordered products onto
\SCETb gives the leading ${\cal O}(\eta^6)$ operators.\footnote{Recall that to derive
  the $\eta^6$, we note that $\lambda^8=\eta^4$, and changing the scaling
  $\lambda\to \eta$ for four collinear quark fields in matching \SCETa $\to$
  \SCETb gives the extra $\eta^2$.  The $\lambda^7$ term gains an extra
  $\lambda$ from the change in scaling to a collinear $D_\perp$.} When combined
with the $\eta^{-7/2}$ from the states this yields a matrix element of order
$\eta^{5/2}$, in agreement with the prefactor in Eq.~(\ref{Aexpn}).  Examples of
the weak operators in \SCETa are
\begin{align} \label{Q01}
  Q_{1d}^{(0)} &=  \big[ \bar u_{n,\omega_1} \bnslash P_L b_v\big]
  \big[ \bar d_{\bn,\omega_2}  \nslash P_L u_{\bn,\omega_3} \big]
  \,,  \nn\\
  Q_{1d}^{(1)} &=
     \big[ \bar u_{n,\omega_1} ig\,\slash\!\!\!\!{\cal B}^\perp_{n,\omega_4} 
     P_L b_v\big]
     \big[ \bar d_{\bn,\omega_2}  \nslash P_L u_{\bn,\omega_3} \big] 
     \,,
\end{align}
where other $Q_{id}^{(0,1)}$ have different flavor structures.  The ``quark''
fields with subscripts $n$ and $\bn$ contain a collinear quark field and
Wilson line with large momenta labels $\omega_i$, such as
\begin{equation}
\bar u_{n,\omega} = \big[ \bar\xi_n^{(u)} W_n\, \delta(\omega\!-\!
  \bn\cdot\cP^\dagger) \big] \,.
\end{equation}
Here $\bar\xi_n$ creates a $n$-collinear quark, or annihilates an antiquark, 
$W_n=W[\bn\cdot A_n]$ is the standard SCET collinear Wilson line built from the
$\bn$ component of $n$-collinear gluons, $\bn\cdot \cP^\dagger$ is an operator
that picks out the large $\bn\cdot p$ label momentum of the fields it acts
on~\cite{SCET}, and $ig\,{\cal B}^{\perp\,\mu}_{n,\omega} = \big[ 1/\bnP\,
W^\dagger_n [ i\bn\mcdot D_{c,n} , i D^\mu_{n,\perp} ] W_n
\delta(\omega-\bnP^\dagger) \big]$. The $b_v$ is an HQET $b$-quark field.

The leading order factorization theorem from \SCETa is~\cite{Bauer:2004tj}
\begin{equation} \label{A0fact}
A^{(0)} = \frac{G_Fm_B^2 f_{M_1}}{\sqrt{2}}\bigg[\! \int_0^1\!\!\! du\, dz\,
  T_{1\!J}(u,z) \zeta^{BM_2}_{J}(z) \phi^{M_1}(u) 
  +\!  \int_0^1\!\!\! du\, T_{1\!\zeta}(u) \zeta^{BM_2} \phi^{M_1}(u) 
 \bigg]\!  + \big\{M_1 \leftrightarrow M_2\big\}.
\end{equation}
Here $T_{1J}$ and $T_{1\zeta}$ contain contributions from the hard scales $m_b$,
and $\phi^M$ is the nonperturbative twist-2 light-cone distribution function.
The terms $\zeta^{BM}$ and $\zeta^{BM}_J(z)$ contain contributions from both the
intermediate scale $\mu_i\sim\sqrt{m_b\Lambda}$ and the scale $\Lambda$, and are
defined by \SCETa matrix elements between $B$ and $M$ states.  In particular
their scaling is
\begin{equation} \label{zeta}
\zeta^{BM}(E),\, \ \zeta_J^{BM}(z,E) \sim
  \bigg(\frac{\Lambda}{m_b}\bigg)^{\!3/2}\, \big[\alpha_s(\mu_i) +\ldots \big],
\end{equation}
explaining the $\alpha_s(\mu_i)$ entry in the $A^{(0)}$ rows of
Table~\ref{table0}. The $\zeta^{BM}$ functions occur in both semileptonic decays
and nonleptonic decays ($E\approx m_B/2$).  Integrating out the scale
$\sqrt{m_b\Lambda}$ to all orders in $\alpha_s$ by matching onto \SCETb
gives~\cite{Bauer:2004tj,Manohar:2006nz}
\begin{align} \label{zetaall}
\zeta_J^{BM}(z,E) &= \frac{f_B f_M m_B}{4E^2}  \int\! dx\! 
  \int\! dk^+\, J(z,x,k_+,E)\, \phi^{M}(x)\, \phi^B_+(k^+) \,, \nn\\
\zeta^{BM}(E) &= \frac{f_B f_M m_B}{4E^2} \sum_{a,b} \int\! dx_1 dx_2\! \int\! dk_1^+
  dk_2^+\,  J_{ab}(x_i,k_j^+,E)\, \phi_a^M(x_i)\, \phi^B_b(k_j^+)\,,
\end{align}
where the $\phi^M_a$ and $\phi^B_b$'s are twist-2 and twist-3, two and three
parton distributions and we pulled out $f_B f_M$ for convenience. The jet
functions $J$, $J_{ab}$ occur due to the time-ordered product structure in
\SCETa and contain contributions from the scale $\sqrt{m_b\Lambda}$.  Using the
result for $\zeta_J^{BM}$ at order $\alpha_s(\mu_i)$ this result agrees with
Ref.~\cite{BBNS} (where expressing $\zeta^{BM}$ in terms of the full theory form
factor generates an additional $\zeta_J^{BM}$ term).  The result for
$\zeta^{BM}$ is from Ref.~\cite{Manohar:2006nz} and required the
MS factorization with zero-bin subtractions. The set of contributing functions
(indices $a,b$) is determined by the complete set of \SCETb operators derived in
Ref.~\cite{LN}. The power counting in $\alpha_s(\mu_i)$ for the \SCETa functions
$\zeta^{BM}$ and $\zeta_J^{BM}$ agree with that derived in
pQCD~\cite{Li:2003yj}.

\begin{table}[t]
\begin{tabular}{|c|c|cc|c|c|}
\hline\hline
Order in  & Time-ordered products 
  & \multicolumn{2}{c|}{Perturbative order}  & Dependence  & 
\raisebox{-6pt}[0pt][-6pt]{Properties}
\\[-4pt]
$\Lambda/m_b$  & in \SCETa  & Annihilation  &  Other & in \SCETb & \\
\hline\hline
$A^{(0)}$ &  $Q_i^{(0)} {\cal L}_{\xi q}^{(1)}$, \ $Q_i^{(0)} {\cal L}_{\xi
   q}^{(2)}$, \ $Q_i^{(0)} {\cal L}_{\xi q}^{(1)} {\cal L}^{(1)}$
    & {\bf ---} &
    $\alpha_s(\mu_i)$  & $\phi_i^{B} \phi_j^{M} \phi^{M'}$
    & Real
  \\
    &  $Q_i^{(1)} {\cal L}_{\xi q}^{(1)}$ 
   & {\bf ---} &
   $\alpha_s(\mu_i)$ & $\phi_+^{B} \phi^{M} \phi^{M'}$
   & Real  \\
  \hline
%
%
$A^{(1)}$ & $Q_i^{(j'=0,1)} {\cal L}_{\xi q}^{(j\le 4)}\, \Pi_{i}\, {\cal 
L}^{(k_i)}$
    & {\bf ---} & $ \alpha_s(\mu_i)$
    &  & Complex
  \\
    & $Q_i^{(4)} $
    & $ \alpha_s(\mu_h)$ & {\bf ---} 
    &  $f_B\, \phi^M \phi^{M'}$ 
    & Real 
  \\
     & $Q_i^{(2)} {\cal L}_{\xi q}^{(1)}$ 
   & $\alpha_s(\mu_h)$  &  $\alpha_s(\mu_i)$
   & $\phi^B\phi^{3M} \phi^{M'}$
   &  Real
   \\
  & $Q_i^{(0)} \big[ {\cal L}_{\xi q}^{(1)} \big]^3 $, \ 
    $Q_i^{(0)} \big[ {\cal L}_{\xi q}^{(1)} \big]^3 {\cal L}^{(1)}$
    &  ${\alpha_s^2(\mu_i)}/{\pi}$  &  $\alpha_s^2(\mu_i)/{\pi}$ 
    &  $S_j(k_{1,2}^+,k_3^-)\ldots $
    & Complex 
  \\
    & $Q_i^{(0)} \big[ {\cal L}_{\xi q}^{(1)} \big]^2 {\cal L}_{\xi
     q}^{(2)} $, \  $Q_i^{(1)} \big[ {\cal L}_{\xi q}^{(1)} \big]^3 $
     &  $\alpha_s^2(\mu_i)/{\pi}$  &  $\alpha_s^2(\mu_i)/{\pi}$ 
    &  $S_j(k_{1,2}^+,k_3^-)\ldots $ 
    & Complex
  \\
      &  $Q_i^{(2)} \big[ {\cal L}_{\xi q}^{(1)} \big]^2 $ & 
   {\bf ---} & $\alpha_s^2(\mu_i)/{\pi}$ & 
   &  Complex
   \\
     & $Q_i^{(2)} {\cal L}_{\xi q}^{(1)} {\cal L}^{(1)}$, 
       $Q_i^{(2)} {\cal L}_{\xi q}^{(2)}$, 
       $Q_i^{(3)} {\cal L}_{\xi q}^{(1)}$
   & $\alpha_s(\mu_h)\, \alpha_s(\mu_i)/{\pi}$  &  $\alpha_s(\mu_i)$
   &
   &  Complex
   \\
  \hline
   $A^{(2)}$  & $Q_i^{(5)} $
    & $ \alpha_s(\mu_h)$ & {\bf ---} 
    &  $ f_B\, \mu_M \phi_{pp}^M \phi^{M'}$ 
    & Real   \\
\hline\hline
\end{tabular}
\caption{All contributions to $B\to M_1 M_2$ amplitudes at leading order
($A^{(0)}$) and at order $\Lambda/m_b$ ($A^{(1)}$), besides $A_{c\bar c}$. In
the first $A^{(1)}$ line $j'+j+\sum k_i \le 4$. The terms with {\bf ---} are
absent or higher order when matched onto \SCETb. The dependence in \SCETb column
lists the known dependence on nonperturbative parameters. The properties column
shows whether at least one of the nonperturbative parameters  is complex. For
$A^{(2)}$, suppressed by $\Lambda^2/m_b^2$, only the local chirally enhanced
annihilation operator is shown.}
\label{table0}
\end{table}

Next we classify the contributions to the power suppressed $B\to M_1 M_2$
amplitudes $A^{(1)}$.  In \SCETa we need to study operators and time-ordered
products with scaling up to ${\cal O}(\lambda^{10})$. These have one or three
light usoft quark fields. The relevant terms are listed in Table~\ref{table0},
where $Q_i^{(j)}\sim \lambda^{6+j}$ and our notation for the Lagrangians up to
second order is taken from Ref.~\cite{bps5}. All the listed terms have an odd
number of soft light quark fields. A basis for the $Q_i^{(4)}$ operators is
constructed in section~\ref{sec:local}, for the $Q_i^{(2)}{\cal L}_{\xi
  q}^{(1)}$ terms in Ref.~\cite{Arnesen2}, and for the $Q_i^{(5)}$ terms in
section~\ref{sec:chiral}.  A basis is not yet known for the remaining $Q_i^{(2)}$
operators, for $Q_i^{(3)}$, and for the ${\cal L}_{\xi q}^{(3,4)}$ and ${\cal
  L}^{(3)}$ Lagrangians, but they do not contribute at ${\cal O}(\alpha_s)$, and
only general properties of these operators are required for our analysis.
Dashes in Table~\ref{table0} indicate terms that are absent to all orders in
$\alpha_s$ for reasons to be explained below.  To determine the perturbative
order listed in the table we count the number of hard $\alpha_s(\mu_h)$ factors
from the matching onto \SCETa, and the number of intermediate scale
$\alpha_s(\mu_i)$ factors from matching onto \SCETb.  The dependence in \SCETb
column lists the nonperturbative quantities that appear in the factorization
theorem for the leading order result described above, and from the factorization
theorems we will derive in sections~\ref{sec:local} and~\ref{sec:chiral} below.
The properties column lists whether the nonperturbative distribution functions
are complex or real as described in detail in section~\ref{sec:phase}, and has
implications for strong phase information in the power corrections.  The results
in Table~\ref{table0} imply the following power counting (for amplitudes not
involving $A_{c\bar c}$),
\begin{align}
{\rm Re}\big[ \hat A^{(0)} \big] & \sim \alpha_s(\mu_i) \,,
  & {\rm Im}\big[ \hat A^{(0)} \big] & \sim \alpha_s(\mu_i)\, \alpha_s(\mu_h) 
  \,,\nn \\
{\rm Re}\big[ \hat A_{ann}^{(1)} \big] & \sim \big[
  \alpha_s(\mu_h)\, + \alpha_s^2(\mu_i) \big]\, \frac{\Lambda}{m_b} \,,
  & {\rm Im}\big[ \hat A_{ann}^{(1)} \big] & \sim \alpha_s^2(\mu_i)\,
  \frac{\Lambda}{m_b}\,, \nn\\
{\rm Re}\big[ \hat A_{rest}^{(1)} \big] & \sim
  \alpha_s(\mu_i)\, \frac{\Lambda}{m_b}\,,
  & {\rm Im}\big[ \hat A_{rest}^{(1)} \big] & \sim \alpha_s(\mu_i)\,
  \frac{\Lambda}{m_b}\,.
\end{align}
To facilitate the discussion we divide the annihilation amplitudes into local
annihilation contributions, $A^{(1,2)}_{Lann}$ from the operators $Q_i^{(4,5)}$
that are insensitive to the jet scale, and into the remaining annihilation
amplitudes, $A^{(1)}_{Tann}$, which are from time-ordered products in \SCETa.
Thus,
\begin{equation}
  A^{(1)}_{ann} = A^{(1)}_{Lann} + A^{(1)}_{Tann} \,.
\end{equation}

\begin{figure}
  \centerline{ 
   \includegraphics[width=6.4cm]{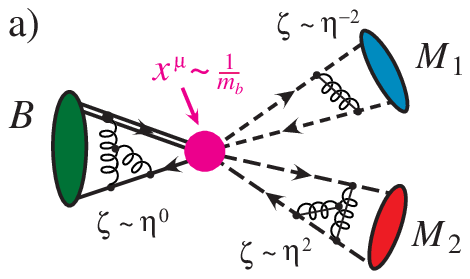} \qquad\quad
   \includegraphics[width=6.6cm]{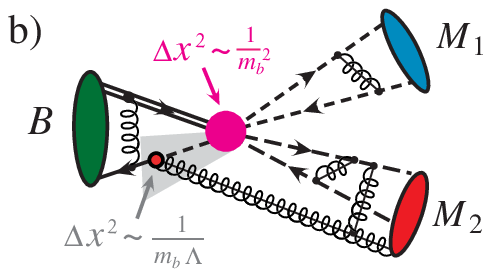}
  } \vspace{0.4cm}
  \centerline{ 
   \includegraphics[width=7cm]{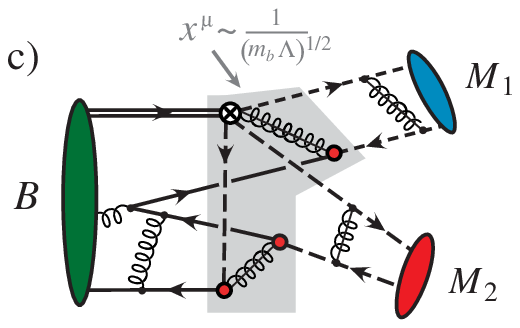} 
  } \label{fig:Ann_FACT}
\caption{Three types of factorization contributions to annihilation amplitudes
  which are the same order in $\eta=\Lambda_{\rm QCD}/m_b$.  a) shows
  $Q_i^{(4)}$ which has $\ge 1$ hard gluon and factorizes at the scale $m_b$.
  The rapidity parameter, $\zeta= p^-/p^+$, controls the MS-factorization between
  soft momenta ($B$), $n$-collinear momenta ($M_2$), and $\bn$-collinear momenta
  ($M_1$).  b) shows the time-ordered product $Q_i^{(2)}{\cal L}^{(1)}_{\xi q}$,
  which involves factorization at $m_b$ and $\sqrt{m_b\Lambda}$. c) shows the time-ordered product $Q_i^{(1)} [{\cal
  L}^{(1)}_{\xi q}]^3$, which factorizes at the scale $\sqrt{m_b\Lambda}$ and
  does not need a hard gluon.  Graphs a) and b) are of order 
  $\alpha_s(\mu_h)$, while c) is $\alpha_s(\mu_i)^2$.}
\end{figure}

In the literature~\cite{Keum,Keum:2000wi,BBNS2,BN,alex} only local annihilation
amplitudes have been studied, and their matrix elements were parameterized
by complex amplitudes. In SCET, $Q_i^{(4)}$ is a six-quark operator with one
usoft quark, such as
\begin{equation} \label{preOp}
   \big( \bar d_s \Gamma_s b_v \big)
   \big( \bar u_{\bn,\omega_2} \Gamma_\bn q_{\bn,\omega_3} \big) 
   \big( \bar q_{n,\omega_1} \Gamma_n u_{n,\omega_4} \big) \,,
\end{equation}
where other $Q_i^{(4)}$ operators have different flavor structures.  To derive
the power counting for this operator, start with $Q^{(0)}\sim \lambda^6$, then
note that switching a collinear quark to an usoft quark costs $\lambda^2$, and
adding a $\xi_n$ and $\xi_\bn$ from a hard gluon also costs $\lambda^2$. This
yields $Q_i^{(4)}\sim {\cal O}(\alpha_s(\mu_h) \lambda^{10})$.  In matching onto
\SCETb we simply replace $Q_i^{(4)}\to O_i^{(1L)}\sim \eta^7$, with the operator
having an identical form. \SCETa operators $Q_i^{(4)}$ that do not have the form
in Eq.~(\ref{preOp}) exist, but they must be taken in time-ordered products
with a subleading Lagrangian and so do not contribute to $A^{(1)}$. For this
reason all local operator contributions to $A^{(1)}$ contribute in the
annihilation terms and not in $A^{(1)}_{rest}$.  Since the matching onto
$O_i^{(1L)}$ is local, it appears as in Fig.~\ref{fig:Ann_FACT}a with an
$\alpha_s(\mu_h)$, but with no jet function.  Thus this contribution to
$A_{ann}^{(1)}$ is of order $\alpha_s(\mu_h)/\alpha_s(\mu_i)\, \Lambda/m_b$
relative to $A^{(0)}$. In section~\ref{sec:local} we construct a complete basis
of $Q^{(4)}_i$ operators and show that their matrix elements are factorizable in
SCET at any order in perturbation theory, and do not generate strong phases at
${\cal O}(\alpha_s(\mu_h))$. We prove a similar theorem for chirally enhanced
terms in the set $Q^{(5)}_i$ in section~\ref{sec:chiral}.

The annihilation amplitudes and other $\Lambda/m_b$ suppressed amplitudes also
occur through time-ordered products. Two examples are shown by
Figs.~\ref{fig:Ann_FACT}b and \ref{fig:Ann_FACT}c. A subset of these terms were
considered in Ref.~\cite{Feldmann:2004mg}, including the diagram in
Fig.~\ref{fig:Ann_FACT}c, and the phenomenological impact of these power
corrections was studied.  So far no attempt has been made to work out the strong
phase properties and perturbative orders in $\alpha_s$ of the time-ordered
products, a task we take up here.  A complete classification of time-ordered
products for the leading power corrections to $B\to M_1M_2$ is listed in
Table~\ref{table0}.  A subset of these terms contribute to the annihilation
amplitudes.  To see which, we note that terms with a $Q^{(0,1)}_i$ and only one
${\cal L}_{\xi q}^{(1)}$ do not contribute to annihilation at either leading or
next-to-leading order; the weak operator is not high enough order in $\lambda$
to contain an extra $n$--$\bn$ pair, and there are not enough ${\cal L}_{\xi
  q}$'s to produce the pair through a soft quark exchange. To rule out these
terms it was important that we are not considering isosinglet final states,
which receive emission annihilation contributions already at leading order. The
term $Q_i^{(2)} [{\cal L}_{\xi q}^{(1)}]^2$ does not contribute to annihilation
because we find that all annihilation type contractions are further power
suppressed when matched onto \SCETb.

Time-ordered products with either a $Q^{(j\ge 2)}$ or with three ${\cal L}_{\xi
  q}$'s do contribute to annihilation. Examples of these two types are shown in
Figs.~\ref{fig:Ann_FACT}b and \ref{fig:Ann_FACT}c. Compared to the local
annihilation amplitude from $Q_i^{(4)}$, only the time-ordered product
$Q^{(2)}{\cal L}_{\xi q}^{(1)}$ contributes at the same order in $\alpha_s$. To
demonstrate this, note that for terms with three ${\cal L}_{\xi q}$'s all graphs
have at least two contracted hard-collinear gluons and so are ${\cal
  O}(\alpha_s^2(\mu_i))$. Graphs with a $Q^{(2,3)}$ start with one
$\alpha_s(\mu_h)$, and will also have an additional $\alpha_s(\mu_i)$ from a
hard collinear gluon, unless it remains uncontracted in matching onto \SCETb.
The uncontracted gluon costs an additional $\lambda$ in the matching onto
\SCETb, so only the time-ordered product $Q^{(2)}{\cal L}_{\xi q}^{(1)}$ can
have a leading, ${\cal O}(\alpha_s(m_b))$, contribution.
Fig.~\ref{fig:Ann_FACT}b gives an example of a diagram occurring from this
time-ordered product. The resulting amplitude involves the three-parton
distribution, $\phi_{3M_2}$. As shown in Ref.~\cite{Arnesen2} it also involves
the twist-2 distribution $\phi_B^+$, and its leading order convolution integrals
converge.

The time-ordered products with three ${\cal L}_{\xi q}$'s are suppressed by
$\alpha_s^2(\mu_i)/\alpha_s(\mu_h)$ relative to $Q^{(4)}_i$, and can be proven
to involve a complex nonperturbative function, as labeled in Table~\ref{table0}
(an example is shown in Fig.~\ref{fig:Ann_FACT}c).  Thus, if perturbation theory
converges rapidly at the scale $\mu_i$, then complex annihilation amplitudes are
highly suppressed. If perturbation theory at $\mu_i$ is poorly convergent then
the time-ordered product contribution could be important numerically; comparable
or even larger than the leading local annihilation amplitude from $Q_i^{(4)}$.
Local annihilation contributions are discussed in detail in
sections~\ref{sec:local} and~\ref{sec:chiral}, while strong phase properties of
the amplitudes and the time-ordered product contributions are taken up in
section~\ref{sec:phase}.


\section{\boldmath Local six-quark operators in \SCETb}
\label{sec:local}


In this section we construct a complete basis of $O_i^{(1L)}$ operators in
\SCETb (the $Q_i^{(4)}$ terms in \SCETa) and derive a factorization theorem for
their contributions to $B\to M_1M_2$.  To find a complete basis we consider
color, spin, and flavor structures that could appear when matching at any order
in $\alpha_s$.  Color is simple, the six-quark operator must have $\Gamma_s
\otimes \Gamma_\bn \otimes \Gamma_n = 1\otimes 1 \otimes 1$.  Although operators
with a $T^A$ in one or more bilinears are allowed at this order, with the
factorization properties of the leading Lagrangians and $\langle M_{n} M_\bn | O
| B_s \rangle = \langle 0 | \ldots | B_s\rangle \langle M_\bn | \ldots | 0
\rangle \langle M_n | \ldots | 0 \rangle $, the terms with $T^A$'s give
vanishing matrix element between the color singlet hadronic
states~\cite{Bauer:2001cu}.

For spin we start by looking at chirality which is preserved by the matching at
$m_b$. Since there is no jet function, the soft spectator quark that
interpolates for the $B$-meson must come from the original operator in $H_W$,
and we Fierz this $\bar\psi$ field next to the $b$-quark field. To be definite,
we take the other $\bar\psi$ field from $H_W$ to go in the $\bn$ direction (in
the SCET Hamiltonian we sum over $n\leftrightarrow \bn$). This implies that the
pair-produced quark is in the $n$ direction. For $O_{1-4,9,10}$ the allowed
chiral structures induced in SCET by matching are $(LH)(LL)(LL)$ and
$(LH)(LR)(RL)$ where $L$ and $R$ correspond to the handedness for the light
quarks in the bilinears in the order shown in Eq.~(\ref{preOp}).  We cannot
assign a handedness to the heavy quark denoted here by $H$. For $O_{5-8}$ we can
have $(LH)(RL)(LR)$, $(LH)(RR)(RR)$, $(RH)(LL)(LR)$, and $(RH)(LR)(RR)$.  A
complete basis of Dirac structures for the individual bilinears is:
\begin{equation} \label{basis}
  \Gamma_s =\gamma^\alpha \,, \qquad
  \Gamma_\bn = \{ \nslash, \nslash\gamma_\perp^\nu \} \,, \qquad
  \Gamma_n = \{ \bnslash, \bnslash\gamma_\perp^\mu \} \,. 
\end{equation}
Structures with $\gamma_5$ are not needed because we have specified the
handedness. Here $\bnslash\gamma_\perp^\mu$ and $\nslash\gamma_\perp^\nu$
connect left and right-handed quarks, while $\bnslash$ and $\nslash$ connect
quarks of the same handedness. From the basis in Eq.~(\ref{basis}) we must
construct an overall scalar using the tensors $v^\mu$, $n^\mu$, $\bn^\mu$,
$g_\perp^{\mu\nu}$, and $\epsilon_\perp^{\mu\nu}\equiv
\epsilon^{\mu\nu\alpha\beta} \bn_\alpha n_\beta/2$. We take $\epsilon^{0123}=1$,
and work in a frame where $v_\perp^\mu=0$ and $n\cdot v =\bn\cdot v =1$, which
makes the set $\{n,\bn,v\}$ redundant. For reasons that will become apparent we
pick $v^\mu$ and $(n^\mu - \bn^\mu)$ as our basis in this section.  The definite
handedness allows us to turn any contraction involving
$i\epsilon_\perp^{\mu\nu}$ into a contraction with $g_\perp^{\mu\nu}$, for
example $i\epsilon_\perp^{\mu\nu} \bar \xi_n^L \bnslash \gamma^\perp_\nu \xi_n^R
= \bar \xi_n^L \bnslash \gamma_\perp^\mu\gamma_5 \xi_n^R = \bar \xi_n^L \bnslash
\gamma_\perp^\mu \xi_n^R$. The $(LH)(LR)(RL)$ and $(LH)(RL)(LR)$ structures can
be ruled out since
\begin{equation}\label{vanish}
  \nslash \gamma_\perp^\mu P_{R} \otimes \bnslash \gamma^\perp_\mu P_{L} 
  = \nslash \gamma_\perp^\mu P_{L} \otimes \bnslash \gamma^\perp_\mu P_{R}
  = 0 \,.
\end{equation}
Noting that $\vslash h_v =h_v$ this leaves four allowed spin structures
\begin{equation} \label{dirac}
  \Gamma_s \otimes \Gamma_\bn \otimes \Gamma_n 
   = \big\{ 1 \otimes \nslash \otimes \bnslash , \
       ( \bnslash \!-\! \nslash) \otimes \nslash \otimes \bnslash , \
       \gamma_\perp^\alpha \otimes \nslash \otimes \bnslash
       \gamma^\perp_\alpha , \
       \gamma_\perp^\alpha \otimes \nslash \gamma^\perp_\alpha \otimes \bnslash
       \big\} \,.
\end{equation}
The last two structures have $\bar q_s \gamma_\perp^\alpha b_v$ and vanish
identically for $B$-meson decays (they would contribute for $B^*$'s).
Furthermore, the local annihilation operators are not sensitive to the $k^+$
momentum of the soft spectator quark. Thus in taking the matrix element we can use
\begin{equation} \label{Bresult}
  \langle 0 | \bar q_s \gamma_5 h_v | B \rangle = -i m_B f_B \,, \qquad
  \langle 0 | \bar q_s \gamma_5 ( \bnslash \!-\! \nslash) h_v | B \rangle 
   = 0 \,.
\end{equation}
Here $f_B$ is the decay constant in the heavy quark limit. The fact that we can
match onto a basis of local SCET operators of the form in Eq.~(\ref{preOp})
demonstrates to all orders in $\alpha_s$ that the local annihilation
contributions are proportional to $f_B$. Using Eq.~(\ref{Bresult}) the second
Dirac structure in Eq.~(\ref{dirac}) is eliminated, so we do not list operators
with $( \bnslash \!-\! \nslash)$ in the soft bilinears below.

Next we consider the allowed flavor structures. From the operators $O_{1,2}$ we
have $(\bar u b)(\bar d q)(\bar q u)$, $(\bar d b)(\bar u q)(\bar q u)$, from
$O_{1-6,7\gamma,8g}$ we have $(\bar d b)(\bar q' q)(\bar q q')$, $(\bar q'
b)(\bar d q)(\bar q q')$, and $O_{7-10}$ give a combination of these. Here the
$q\bar q$ are the pair produced $n$ and $\bn$ pair, while the $q'\bar q'$
appeared in the weak operators.  Thus a basis for $B$-decay operators is
\begin{eqnarray} 
  O_{1d}^{(1L)} &=&  \frac{2}{m_b^3}\, {\textstyle \sum}_q \,
   \big[ \bar d_{s}P_R   {b}_{v} \big]
  \big[ \bar u_{\bn,\omega_2}  \nslash P_L\, q_{\bn,\omega_3} \big]
  \big[ \bar q_{n,\omega_1}  \bnslash P_L u_{n,\omega_4} \big]
   \,, \nn \\ 
  O_{2d}^{(1L)} &=&  \frac{2}{m_b^3}\, {\textstyle \sum}_q \,
   \big[ \bar u_{s}P_R   {b}_{v} \big]
  \big[ \bar d_{\bn,\omega_2}  \nslash P_L\, q_{\bn,\omega_3} \big]
  \big[ \bar q_{n,\omega_1}  \bnslash P_L u_{n,\omega_4} \big]
   \,, \nn \\
  O_{3d}^{(1L)} &=& \frac{2}{m_b^3}\, {\textstyle \sum}_{q,q'} \,
   \big[ \bar d_{s}P_R   {b}_{v} \big]
  \big[ \bar q'_{\bn,\omega_2}  \nslash P_L\, q_{\bn,\omega_3} \big]
  \big[ \bar q_{n,\omega_1}  \bnslash P_L q'_{n,\omega_4} \big]
   \,, \nn \\ 
  O_{4d}^{(1L)} &=& \frac{2}{m_b^3}\, {\textstyle \sum}_{q,q'}  \,
   \big[ \bar q'_{s}P_R   {b}_{v} \big]
  \big[ \bar d_{\bn,\omega_2}  \nslash P_L\, q_{\bn,\omega_3} \big]
  \big[ \bar q_{n,\omega_1}  \bnslash P_L q'_{n,\omega_4} \big]
   \,, \nn \\ 
  O_{5d}^{(1L)} &=&  \frac{2}{m_b^3}\, {\textstyle \sum}_{q}  \,
   \big[ \bar d_{s} P_R  {b}_{v} \big]
  \big[ \bar u_{\bn,\omega_2}  \nslash P_R\, q_{\bn,\omega_3} \big]
  \big[ \bar q_{n,\omega_1}  \bnslash P_R\, u_{n,\omega_4} \big]
   \,, \nn \\ 
  O_{6d}^{(1L)} &=& \frac{2}{m_b^3}\, {\textstyle \sum}_{q}  \,
   \big[ \bar u_{s} P_R  {b}_{v} \big]
  \big[ \bar d_{\bn,\omega_2}  \nslash P_R\, q_{\bn,\omega_3} \big]
  \big[ \bar q_{n,\omega_1}  \bnslash P_R\, u_{n,\omega_4} \big]
   \,, \nn 
\end{eqnarray}
\begin{eqnarray} \label{Qann}
   O_{7d}^{(1L)} &=&  \frac{2}{m_b^3}\, {\textstyle \sum}_{q,q'}  \,
   \big[ \bar d_{s} P_R  {b}_{v} \big]
  \big[ \bar q'_{\bn,\omega_2}  \nslash P_R\, q_{\bn,\omega_3} \big]
  \big[ \bar q_{n,\omega_1}  \bnslash P_R\, q'_{n,\omega_4} \big]
   \,, \nn \\ 
  O_{8d}^{(1L)} &=& \frac{2}{m_b^3}\, {\textstyle \sum}_{q,q'}  \,
   \big[ \bar q'_{s} P_R  {b}_{v} \big]
  \big[ \bar d_{\bn,\omega_2}  \nslash P_R\, q_{\bn,\omega_3} \big]
  \big[ \bar q_{n,\omega_1}  \bnslash P_R\, q'_{n,\omega_4} \big]
   \,. 
\end{eqnarray}
Here we integrated out $c$ and $b$ quarks in the sum over flavors, so the
remaining sums are over $q=u,d,s$ and $q'=u,d,s$.  For the $\Delta S=0$
effective Hamiltonian with Wilson coefficients $a_i^{(d)}(\omega_j)$ we use the
notation
\begin{equation} \label{match}
 H_W =  \frac{4G_F}{\sqrt{2}}\, \sum_{n,\bn} 
   \int [d\omega_{1}d\omega_{2}d\omega_{3}  d\omega_{4}]\, 
    \sum_{i=1-8} a_i^{d}(\omega_j)\,  O_{id}^{(1L)}(\omega_j) \,.
\end{equation}
To pull the CKM structures out of the SCET Wilson coefficients we write
\begin{equation}
a_i^{d}(\omega_j) = \Bigg\{ 
  \begin{array}{lc}
  \lambda^{(d)}_u \, a_{iu}(\omega_j) 
    + \lambda^{(d)}_c \, a_{ic}(\omega_j) 
 \qquad & 
   [i=1,2,3,4]\,, \\[4pt]
   (\lambda^{(d)}_u + \lambda^{(d)}_c) \, a_{i}(\omega_j) 
  &
   [i=5,6,7,8]\,,
  \end{array}
\end{equation}
where $\lambda_p^{(d)}= V_{pb} V^*_{pd}$. Identical definitions for $a_i^{s}$
are made by replacing $\lambda_u^{(d)}\to \lambda_u^{(s)}$ and
$\lambda_c^{(d)}\to \lambda_c^{(s)}$. For $i=5,6,7,8$ only penguin operators
contribute.

Next we take the $B\to M_1M_2$ matrix element of $H_W$. The factorization
properties of SCET yield
\begin{align} \label{factor}
  \langle M_1 M_2 | O_{1d}^{(1L)} |B \rangle &= \frac{2}{m_b^3}\, 
   {\textstyle \sum}_q \,
   \langle M_1 | \bar u_{\bn,\omega_2}  \nslash P_L\, q_{\bn,\omega_3} 
   | 0 \rangle
   \langle M_2 |  \bar q_{n,\omega_1}  \bnslash P_L u_{n,\omega_4} 
   | 0 \rangle
   \langle 0 |  \bar d_{s}P_R   {b}_{v} | B \rangle  \nn\\
 & \quad + \big\{M_1 \leftrightarrow M_2\big\} \,,
\end{align}
with similar results for the other $O_{id}^{(1L)}$ terms. Here the
$\{M_1\leftrightarrow M_2\}$ indicates terms where the flavor
quantum numbers of the $M_2$ state match those of the $\bn$-collinear operator.
The matrix elements in Eq.~(\ref{factor}) are zero for transversely polarized
vector mesons in agreement with the helicity counting in Ref.~\cite{alex}.
Equation~(\ref{factor}) can be evaluated using Eq.~(\ref{Bresult}) and
\begin{align} \label{lcdistn}
  \langle P_{n_1}(p) | \bar q^{(f)}_{n,\omega}\, \bnslash P_{L,R}\,
  q^{(f')}_{n,\omega'} | 0 \rangle 
  &=  \frac{\pm i\, f_P }{2}\, c_{Pff'}\, \delta_{nn_1}\,
   \delta(\bn\mcdot p \!-\! \omega\!+\!
  \omega')\, \phi_P(y) \,, \nn\\
 \langle V_{n_1}(p,\varepsilon) | 
   \bar q^{(f)}_{n,\omega}\, \bnslash P_{L,R}\, 
  q^{(f')}_{n,\omega'} | 0 \rangle &= \frac{i f_V  m_V {\bn\mcdot
      \varepsilon}}{2\,\bn\mcdot p}\, c_{Vff'}\, \delta_{nn_1}\,
   \delta(\bn\mcdot p \!-\! \omega\!+\! \omega')\, \phi_{V_\parallel}(y)
  \,.
\end{align}
Here $f,f'$ are flavor indices, $\phi_P(y)$ and $\phi_{V_\parallel}(y)$ are the
twist-2 light-cone distribution functions for pseudoscalars and vectors,
$y=\omega/\bn\cdot p = \omega/m_b$, and $c_{Pff'}$, $c_{Vff'}$ are
Clebsch-Gordan coefficients. For the $M_2$ mesons, $P_{n_2}$ and $V_{n_2}$, we
have the same equation with $n\leftrightarrow \bn$, and $y\to x$. Since the
$P_{L,R}$ only induce $\pm$ signs in the pseudoscalar matrix element, it is
convenient to define
\begin{table}[t]
\begin{tabular}{|c|c|}
\hline\hline
$M_1 M_2$  & $H(x,y)$  
\\ \hline\hline 
$\pi^- \pi^+$,  $\pi^-\rho^+$, $\rho^-\pi^+$, $\rho^- \rho^+$ & 
 $-\tilde a^d_1(x,y)-\tilde a^d_4(y,x) -  \tilde a^d_3(x,y)  - \tilde
 a_3^d(y,x)$
 \\
  $\pi^- \pi^0$, $\rho^-\pi^0$ $\pi^- \rho^0$, $\rho^-_\parallel \rho^0_\parallel$& 
   $\frac{1}{\sqrt2} \big[\tilde a^d_2(x,y) + \tilde a^d_4(x,y)
     - \tilde a^d_2(y,x) - \tilde a^d_4(y,x) \big] 
    $ \\
$\pi^0 \pi^0,$ $\pi^0 \rho^0$,  $\rho^0 \rho^0$ &  
  $ \big[ \frac12 \tilde a^d_1(x,y)+ \tilde a^d_3(x,y)+ \frac12 \tilde a^d_4(x,y) \big]
  + \big[ x\leftrightarrow y \big] $ 
  \\
 $K^{(*)-} K^{(*)+}$ &  
  $- \tilde a^d_1(x,y) - \tilde a^d_3(x,y) - \tilde a^d_3(y,x) $ 
  \\
  $\bar K^{(*)0} K^{(*)0}$ 
  & $ \tilde a^d_3(x,y)  + \tilde a^d_3(y,x) + \tilde a^d_4(x,y)$ 
  \\
 $K^{(*)-} K^{(*)0}$ & $\tilde a^d_2(x,y)+ \tilde a^d_4(x,y)$
  \\
\hline 
 $\pi^- \bar K^{(*)0}$, $\rho^- \bar{K}^{(*)0}$ 
   &$ \tilde a_2^s(x,y) + \tilde a_4^s(x,y) $
  \\
 $\pi^0 \bar{K}^{(*)-},\rho^0 {K}^{(*)-}$  
   & $-\frac{1}{\sqrt 2}\big[ \tilde a_2^s(x,y)+ \tilde a_4^s(x,y) \big] $ 
   \\
 $\pi^0 \bar K ^{(*)0},\rho^0 \bar K ^{(*)0}$ 
   & $ \frac{1}{\sqrt 2} \, \tilde a_4^s(x,y) $
   \\
 $\pi^+ K^{(*)-} ,\rho^+ K^{(*)-}$ 
   &  $- \tilde a_4^s(x,y)$\\
  \hline\hline
\end{tabular}
\caption{Hard functions for $\bar B^0$ and $B^-$ decays for the annihilation 
amplitude $A_{Lann}^{(1)}$ in Eq.~(\ref{Aann}). For each pair of mesons in 
the table, the first is $M_1$ and the second $M_2$.}
\label{table1a}
\end{table}
\begin{table}[t]
\begin{tabular}{|c|c|}
\hline\hline
$M_1 M_2$  & $H(x,y)$  
\\ \hline\hline 
$\pi^- K^{(*)+}$, $\rho^- K^{(*)+}$ & 
 $-\tilde a^d_4(y,x) $
 \\
 $\pi^0 K^{(*)0}$,  $\rho^0 K^{(*)0}$ &  
  $  \frac{1}{\sqrt 2}\,  \tilde a^d_4(y,x)  $ 
  \\
\hline
$\pi^- \pi^+$,  $\pi^-\rho^+$, $\rho^-\pi^+$, $\rho^- \rho^+$ & 
 $-\tilde a^s_1(x,y) -  \tilde a^s_3(x,y)  - \tilde
 a_3^s(y,x)$
 \\
 $\pi^0 \pi^0,$ $\pi^0 \rho^0$,  $\rho^0 \rho^0$ &  
  $ \big[ \frac12 \tilde a^s_1(x,y)+ \tilde a^s_3(x,y) \big]
  + \big[ x\leftrightarrow y \big] $ 
  \\
 $K^{(*)-} K^{(*)+}$ &  
  $-\tilde a^s_1(x,y)-\tilde a^s_4(y,x) -  \tilde a^s_3(x,y)  - \tilde
 a_3^s(y,x)$
  \\
  $\bar K^{(*)0} K^{(*)0}$ 
  & $ \tilde a^s_3(x,y)  + \tilde a^s_3(y,x) + \tilde a^s_4(y,x)$ 
  \\
  \hline\hline
\end{tabular}
\caption{Hard functions  for $\bar B_s$ decays for the annihilation amplitude
 $A_{Lann}^{(1)}$ in Eq.~(\ref{Aann}).}
\label{table1b}
\end{table}

\begin{equation}
  \tilde a_1^d = a_1^d + \kappa a_5^d \,, \qquad
  \tilde a_2^d = a_2^d + \kappa a_6^d \,, \qquad
  \tilde a_3^d = a_3^d + \kappa a_7^d \,, \qquad
  \tilde a_4^d = a_4^d + \kappa a_8^d \,,
\end{equation}
with similar definitions for $\tilde a_i^s$.  Here $\kappa=+1$ for $PP$, $VV$,
and $\kappa=-1$ for $PV$ channels.  Using these results, the ${\cal
  O}(\Lambda/m_b)$ local annihilation amplitudes are
\begin{equation} \label{Aann}
  A^{(1)}_{Lann}(\bar B\to M_1 M_2) = - \frac{G_F f_B f_{M_1} f_{M_2} }{\sqrt{2}} 
  \int_0^1\!\! dx\, dy\, H(x,y)\, \phi^{M_1}(y) \phi^{M_2}(x) \,.
\end{equation} 
Here $H(x,y)$ are perturbatively calculable hard coefficients determined by the
SCET Wilson coefficients $\tilde a_i(\omega_j)$. Results for different final
states are listed in Table~\ref{table1a} for $\bar B^0$ and $B^-$ decays, and in
Table~\ref{table1b} for $\bar B_s$ decays.  Our derivation of the local
annihilation amplitude in Eq.~(\ref{Aann}) is valid to all orders in $\alpha_s$,
and provides a proof of factorization for this term.

\begin{figure}[t]
\includegraphics[width=16cm]{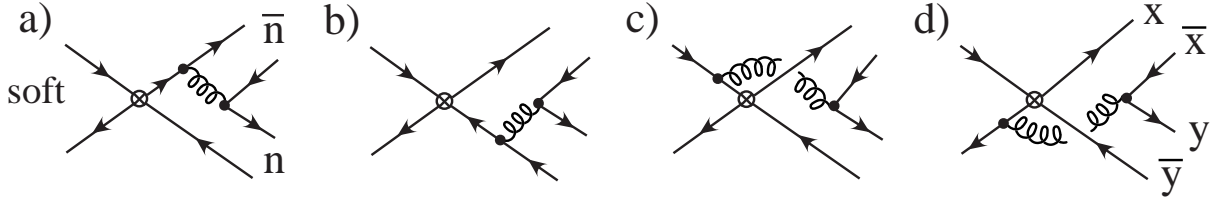}
\caption{Tree level annihilation graphs for $B\to M_1 M_2$ decays. Here soft,
  $n$, $\bn$ denote quarks that are soft, $n$-collinear, and $\bn$-collinear 
  respectively.
  \label{fig:Ann}}
\end{figure}

Matching at tree level, involves computing the ${\cal O}(\alpha_s(m_b))$ graphs
in Fig.~\ref{fig:Ann} and comparing them with matrix elements of the SCET
operators $Q_i^{(4)}$. Doing so we find that the Wilson coefficients $a_i(x,y)$
are
\begin{align} \label{aLO}
 a_{1u} &= \frac{C_F \pi \alpha_s(\mu_h) }{N_c^2}\,  F(x,y)\,
    \Big( C_1 +\frac32\, C_{10} \Big)\,,
& a_{1c} &= \frac{C_F \pi \alpha_s(\mu_h) }{N_c^2}\,  F(x,y)\,
    \Big( \frac32\, C_{10} \Big)\,, \nn \\
  a_{2u} &= \frac{C_F \pi \alpha_s(\mu_h) }{N_c^2}\,  F(x,y)\,
    \Big(C_2 + \frac32\, C_9 \Big) \,,
& a_{2c} &= \frac{C_F \pi \alpha_s(\mu_h) }{N_c^2}\,  F(x,y)\,
    \Big(\frac32\, C_9 \Big) \,,
    \nn \\
 a_{3u}&= \frac{C_F \pi \alpha_s(\mu_h) }{N_c^2}\, F(x,y)\,
    \Big(C_4 -\frac12\, C_{10} \Big)\,,
& a_{3c} &= \frac{C_F \pi \alpha_s(\mu_h) }{N_c^2}\, F(x,y)\,
    \Big(C_4 -\frac12\, C_{10}  \Big)\,, \nn \\
a_{4u} &= \frac{C_F \pi \alpha_s(\mu_h) }{N_c^2}\, F(x,y)\,
    \Big(  C_3-\frac12\, C_{9} \Big) \,,
 &a_{4c} &= \frac{C_F \pi \alpha_s(\mu_h) }{N_c^2}\, F(x,y)\,
    \Big(  C_3-\frac12\, C_{9}   \Big) \,,
    \nn \\
 a_5&= \frac{C_F \pi \alpha_s(\mu_h) }{N_c^2}\, F(\bar y,\bar x)
   \Big( \frac32\, C_8 \Big)\,,
 &a_6 &= 0\,,
    \nn \\
 a_7&= \frac{C_F \pi \alpha_s(\mu_h) }{N_c^2}\, F(\bar y,\bar x)
   \Big(C_6- \frac12\, C_8   \Big)\,,
 &a_8 &= 0 \,,
\end{align}
where $\mu_h \sim m_b$, $\bar x=1-x$, $\bar y=1-y$, with quark momentum
fractions $x$ and $y$ as defined in Eq.~(\ref{lcdistn}) and shown in
Fig.~\ref{fig:Ann}. The function $F$ is
\begin{equation} \label{F}
F(x,y) = \bigg[\frac1{\bar x^2 y} - \frac1{y( x \bar y-1)}\bigg]_{\mbox{\o}}
  \ +  \frac{d(\mu_-)\:\delta'(\bar x)}{y} \,,
\end{equation}
where the \o-notation and term involving the Wilson coefficient
$d(\mu_-)$ are discussed below.  The function $F(\bar y,\bar x)$ will
involve $d(\mu_+)$. Note that the coefficients $a_{3u,3c,4u,4c,7,8}$
are polluted in the sense of Ref.~\cite{Bauer:2004tj}, meaning that
${\cal O}(\alpha_s^2)$ matching results proportional to the large
coefficients $C_{1,2}$ could compete numerically. The others are not
polluted: $a_{1u,2u}$ involve $C_{1,2}$ at ${\cal O}(\alpha_s)$, while
$a_{1c,2c,5,6}$ only get contributions from electroweak penguins.  Our
results for the diagrams in Fig.~\ref{fig:Ann} agree with
Refs.~\cite{Keum,BBNS2}. This includes the appearance of the
combinations of momentum fractions in the functions $F(x,y)$ and
$F(\bar y,\bar x)$, up to
\o-distribution and $d$-term. For later convenience we define moment parameters
which convolute the hard coefficients with the meson distributions
\begin{align} \label{beta1}
  \beta_{iu}^{M_1M_2} &=  \int_0^1\!\! dx\, dy\,
   [a_{iu}(x,y) \!+\! \kappa a_{i+4}(x,y)]\, \phi^{M_1}(y) \phi^{M_2}(x) 
   \,, \nn\\
   \beta_{ic}^{M_1M_2} &=  \int_0^1\!\! dx\, dy\,
   [a_{ic}(x,y) \!+\! \kappa a_{i+4}(x,y)]\, \phi^{M_1}(y) \phi^{M_2}(x)
    \,.
\end{align}

In Eq.~(\ref{F}) the subscript $\mbox{\o}$ denotes the fact that singular terms
in convolution integrals are finite in SCET due to the MS-factorization which
involves convolution integrals such as
\begin{equation} \label{sums}
 \sum_{x,\, x'\ne 0} \int\! dx_r\, dx'_r\, \delta(1\!-\! x\!-\! x')\,
  \frac{ \phi_M(x,x',\mu) }{\bar x^2} \,,
\end{equation}
where $x^{(\prime)}$ and $x^{(\prime)}_r$ correspond to label and residual
momenta~\cite{Manohar:2006nz}. Implementing $x\ne 0$ and $x'\ne 0$ in the
MS-factorization scheme requires zero-bin subtractions and divergences in the
rapidity must also be regulated.  The $\delta$-function sets $x'=1-x$, so $x'\ne
0$ enforces $x\ne 1$.  With the usual assumption that $\phi_M(x)$ vanishes at
its endpoints with a power-like fall-off slower than quadratic, only integrals
over $1/\bar x^2$ in $F(x,y)$ and $1/y^2$ in $F(\bar y,\bar x)$ require special
care,
\begin{align}
 & \big\langle \bar x^{-2} \big\rangle^{M} = \! \int_0^1\!\!\! dx\,
  \frac{ \phi_M(x,\mu) }{(\bar x^2)_{\mbox{\o}} } \,,
 & \big\langle y^{-2} \big\rangle^{M} & = \! \int_0^1\!\!\! dy\,
  \frac{ \phi_M(x,\mu) }{(y^2)_{\mbox{\o}} } 
  \,.
\end{align}
The resulting moments $\langle \bar x^{-2} \rangle^M$ and $\langle y^{-2}
\rangle^M$ should be considered hadronic parameters, for which we use the
minimal subtraction scheme. Their value depends on $\mu$ and $\mu_\pm$ and are
scheme dependent beyond the usual $\overline{\rm MS}$ scheme for $\phi_M$. This
can be viewed as a modification of the distribution function, $\phi_M(x,\mu)\to
\phi_M(x,\mu,\mu_-)$, where the $x^{-2}$ moment of $\phi_M(x,\mu,\mu_-)$ converges. In
order to derive a result that makes it easy to find a model for these moments we follow
Ref.~\cite{Manohar:2006nz} and assume there is no interference between the
rapidity renormalization and invariant mass renormalization, which gives
\begin{align} \label{Apif}
& \big\langle \bar x^{-2} \big\rangle^{M} 
  =\! \int_0^1\!\!\! dx\, \frac{ \phi_M(x,\mu) + \bar x \phi'_M(1,\mu)}{\bar x^2}
  - \phi'_M(1,\mu)\, \ln\Big( \frac{\bn\mcdot p_M}{\muminus}\Big) 
  \,,\nn\\
& \big\langle y^{-2} \big\rangle^{M} 
  =\! \int_0^1\!\!\! dy\, \frac{ \phi_M(y,\mu) - y \phi'_M(0,\mu)}{y^2}
  + \phi'_M(0,\mu)\, \ln\Big( \frac{n\mcdot p_M}{\muplus}\Big) 
  \,.
\end{align}
Here $\phi_M'(1)$ is generated by a zero-bin subtraction which avoids
double counting the region where $\bar x\to 0$. When $\bar x\to 0$ the
corresponding outgoing quark becomes soft, and this contribution is
taken into account by a time-ordered product term in Table~\ref{table0}.  To obtain the
renormalized $\langle \bar x^{-2}\rangle^M$ result in Eq.~(\ref{Apif})
requires $1/\epsilon_{\rm UV}$ counterterms which correspond to
operators with the $\bn$-collinear bilinears in Eq.~(\ref{Qann}),
$[\bar u_{\bn,\omega_2} \nslash \gamma_5 q_{\bn,\omega_3}]$ etc.,
which can be written as~\cite{Manohar:2006nz}
\begin{align} \label{Oct}
  O_{ct} &= \frac{\partial}{\partial \omega_3} (\bar\xi_\bn W)_{\omega_2} \nslash\gamma_5
  (W^\dagger \xi_\bn)_{\omega_3} \bigg|_{\omega_3\to 0} \,.
\end{align} 
The matrix element of these terms is taken prior to performing the
partial derivative and the limit $\omega_3\to 0$, and gives
$\phi_M^\prime(1,\mu)$.  These terms do not have a $\omega_3\ne 0$
restriction, and consistency of the renormalization procedure used to
obtain Eq.~(\ref{Apif}) demands that the fields here are
$\bn$-collinear. An analogous set of terms are required for
$\phi_M^\prime(0,\mu)$. These terms are real at any scale, which
follows from the requirements discussed in section~\ref{sec:phase} for
an \SCETb operator to be able to generate a physical strong phase. The
dependences on $\mu_\pm$ in Eq.~(\ref{Apif}) are canceled by the
leading dependences on these scales, $d(\mu_-)=\ln(p_M^-/\mu_-)+ \kappa$ and
$d(\mu_+)=\ln(p_M^+/\mu_+)+\kappa$, which appeared in Eq.~(\ref{F}). Here
$\kappa$ can be fixed by a matching computation. The $d(\mu_\pm)$
correspond to the renormalized coefficients of the $O_{ct}$, and must
be included for consistency at this order~\cite{LMS}.  In the rough
numerical analysis we do later on, we will treat the contributions
from these coefficients as part of the uncertainty.

Note that in deriving the result in Eq.~(\ref{F}) we have dropped $i\epsilon$
factors from the propagators. If these terms were kept, the second term in
$F(x,y)$ would be
\begin{equation}
  \frac{1}{(y+i\epsilon)\, (x\bar y -1 +i\epsilon)} \,.
\end{equation}
The $i\epsilon$'s yield imaginary contributions with $\delta(y)$ and
$\delta(x\bar y-1)$. They contribute for $y=0$ or for $x=\bar y=1$, so these
contributions occur in zero-bins, which are excluded from the convolution
integrals in the factorization theorem we have derived with SCET. The zero-bins
correspond to degrees of freedom that are soft, and including these regions
would induce a double counting, so the correct factorization theorem in QCD does
not include them.  Factors analogous to $x\ne 0$ and $x'\ne 0$ in
Eq.~(\ref{sums}) ensure that there is no contribution to the integral from any
zero-bin momentum, and we find that the $\delta$-function terms give zero. This
remains true for more singular distributions yielding $\delta^{(n)}(x)$, and so
also applies to the first term in $F(x,y)$. Thus it is correct to drop the
$i\epsilon$ factors from the start. This should be compared with the approach in
KLS where the $i\epsilon$ factors generate a strong phase from the tree level
diagrams from a $k_\perp^2$ dependent $\delta$-function. In our derivation any
such $k_\perp^2$ imaginary terms could only occur at higher orders in
$\Lambda/m_b$.

Thus at order $\alpha_s(\mu_h)$ the lowest order annihilation factorization
theorem is determined by the convolutions
\begin{align} \label{Fconv}
 & \int_0^1\! dx\, dy\, F(x,y) \phi^{M_1}(y) \phi^{M_2}(x) \\
    &\qquad 
   = \big\langle \bar x^{-2} \big\rangle^{M_2} \big\langle y^{-1}
    \big\rangle^{M_1} - \big\langle [y(x\bar y-1)]^{-1} \big\rangle^{M_1M_2} 
  + d(\mu_-) \phi_{M_2}'(1)  \big\langle y^{-1} \big\rangle^{M_1} 
   \,, \nn \\
 & \int_0^1\! dx\, dy\, F(\bar y,\bar x) \phi^{M_1}(y) \phi^{M_2}(x)  
   \nn\\
    &\qquad 
     = \big\langle y^{-2} \big\rangle^{M_1} \big\langle \bar x^{-1}
    \big\rangle^{M_2}
    - \big\langle [\bar x(x\bar y-1)]^{-1} \big\rangle^{M_1M_2} 
  - d(\mu_+) \phi_{M_1}'(0)  \big\langle \bar x^{-1}
    \big\rangle^{M_2} 
   \,. \nn
\end{align}
Here we use Eq.~(\ref{Apif}), and
\begin{align} \label{Apif2}
 \langle y^{-1} \rangle^M &= \int_0^1\!\! dy\, \frac{\phi^M(y,\mu)}{y} \,, 
  & \langle f(x,y) \rangle^{M_1M_2} & = \int_0^1\!\! dx \!\int_0^1\!\! dy\,
     f(x,y)\, \phi^{M_1}(y,\mu)\, \phi^{M_2}(x,\mu) .
\end{align}
These results do not have a complex phase because the right-hand side of
Eq.~(\ref{Fconv}) is real.

We have shown that the convolution formula in Eq.~(\ref{Aann}) for the local
contributions $O_i^{(1L)}$ yields a well-defined annihilation amplitude.  At
order $\alpha_s(m_b)$ the result is real, so $A^{(1)}_{Lann}$ is real up to
perturbative corrections.  Order $\alpha_s^2(m_b)$ corrections to the $a_i$ will
produce perturbative strong phases in $A^{(1)}_{Lann}$.  Further discussion on
strong phases is given in section~\ref{sec:phase}, while phenomenological
implications are taken up in section~\ref{sec:phen}.


\section{Chirally Enhanced Local Annihilation Contributions}
\label{sec:chiral}


At order $\alpha_s(\mu_h) \mu_M\Lambda/m_b^2$ there are contributions from
chirally enhanced operators that could compete with the $\alpha_s(\mu_h)
\Lambda/m_b$ terms~\cite{BBNS2}.  In SCET we define these contributions as the
set of \SCETb operators analogous to $O^{(1L)}_i$ but with an extra
$\slash\!\!\!\!\cP_\perp$ between collinear quarks fields. We start by
constructing a complete basis for local operators at this order with a
$\cP_\perp^\beta$, calling them $O^{(2L)}_i$. These operators have the same
color and flavor structures as Eq.~(\ref{Qann}).  The chiral structures induced
from the operators $O_{1-10}$ and the initial basis of Dirac structures shown in
Eq.~(\ref{basis}) are also the same, and allow us to eliminate many
possibilities.

The complete set of Dirac structures from matching the operators $O_{1-4,9,10}$
include
\begin{align} \label{gamma3}
  \Gamma_s \otimes \Gamma_\bn \otimes \Gamma_n \cP^\beta_\perp
  = \big\{ &
  \gamma^\perp_\beta \otimes \nslash \otimes \bnslash \cP^\beta_\perp ,\  
  \gamma_\perp^\alpha \otimes \nslash \gamma^\perp_\alpha
     \otimes \bnslash \gamma^\perp_\beta \cP^\beta_\perp ,\  \nn\\
 & \gamma^\perp_\beta \otimes \nslash \gamma_\perp^\alpha
  \otimes \bnslash \gamma^\perp_\alpha \cP^\beta_\perp ,\ 
  \gamma_\perp^\alpha \otimes \nslash \gamma^\perp_\beta
  \otimes \bnslash \gamma^\perp_\alpha \cP^\beta_\perp 
  \big\} \,,
\end{align}
plus the analogous set $\Gamma_s \otimes \Gamma_\bn \cP^\beta_\perp \otimes
\Gamma_n$.  Our basis does not include operators with $\cP^\dagger_\perp$,
because the mesons $M_i$ have zero $\perp$-momenta, so we can integrate these
terms by parts to put them in the form in Eq.~(\ref{gamma3}). The third term in
Eq.~(\ref{gamma3}) has chiral structure $(LH)(LR)(RL)$ and vanishes by
Eq.~(\ref{vanish}). The terms in Eq.~(\ref{gamma3}) all have $\bar q_s
\gamma_\perp^\mu b_v$, and so do not contribute for $B$-decays. The same holds
if we replace ${\cal P}_\perp^\beta$ by $ig {\cal B}_\perp^\beta$.  Thus, at any
order in perturbation theory the only ${\cal O}(\eta^8)$ local operator
contributions from $O_{1-4,9,10}$ are those with a $D_s^\mu$ in the soft
bilinear.

For $O_{5-8}$ we have the structures in Eq.~(\ref{gamma3}), and when the $q'$
flavor is a soft quark with $P_L\otimes P_R$ Dirac structure from $O_i$ we also
have
\begin{align} \label{gamma2}
  & \Gamma_s \otimes \Gamma_\bn \otimes \Gamma_n \cP^\beta_\perp
   = \big\{ 1 \otimes \nslash \otimes \bnslash\: \slash\!\!\!\!\cP_\perp ,\ 
    1 \otimes \nslash\gamma^\perp_\beta \otimes \bnslash \cP_\perp^\beta \big\} 
   \,,\nn\\
    & \Gamma_s \otimes \Gamma_\bn \cP^\beta_\perp \otimes \Gamma_n 
   = \big\{ 1 \otimes \nslash \cP_\perp^\beta \otimes \bnslash\gamma^\perp_\beta  ,\ 
    1 \otimes \nslash\: \slash\!\!\!\!\cP_\perp \otimes \bnslash\,  \big\} 
   \,,
\end{align}
plus operators with 1 replaced by $\bnslash-\nslash$, which vanish due to
Eq.~(\ref{Bresult}).  The operators in Eq.~(\ref{gamma2}) contribute to
$B$-decays.  In particular, they yield both transverse and longitudinal
polarization in $B\to VV$. A complete basis for the local ${\cal
O}(\eta^8)$ operators with one $\cP_\perp^\beta$ is
\begin{align} \label{Q2ann}
  O_{1d}^{(2L)} &= \frac{1}{m_b^4}\, {\textstyle \sum}_{q,q'} \,
   \big[ \bar q'_{s}P_L   {b}_{v} \big]
  \big[ \bar d_{\bn,\omega_2}  \nslash P_L\, q_{\bn,\omega_3} \big]
  \big[ \bar q_{n,\omega_1}  \bnslash\: \slash\!\!\!\!\cP_\perp \! P_R\,
      q'_{n,\omega_4} \big]
   \,, \nn \\ 
  O_{2d}^{(2L)} &= \frac{1}{m_b^4}\, {\textstyle \sum}_{q,q'}  \,
   \big[ \bar q'_{s}P_L   {b}_{v} \big]
  \big[ \bar d_{\bn,\omega_2}  \nslash\: \slash\!\!\!\!\cP_\perp \! P_R\,
      q_{\bn,\omega_3} \big]
  \big[ \bar q_{n,\omega_1}  \bnslash P_R\, q'_{n,\omega_4} \big]
   \,, \nn \\[5pt]
  O_{3d}^{(2L)} &=  \frac{1}{m_b^4}\, {\textstyle \sum}_{q,q'}  \,
   \big[ \bar q'_{s} P_L  {b}_{v} \big]
  \big[ \bar d_{\bn,\omega_2}  \nslash \gamma^\perp_\beta P_R\,
      q_{\bn,\omega_3} \big]
  \big[ \bar q_{n,\omega_1}  \bnslash P_R\,\cP^\beta_\perp q'_{n,\omega_4} \big]
   \,, \nn \\ 
  O_{4d}^{(2L)} &= \frac{1}{m_b^4}\, {\textstyle \sum}_{q,q'}  \,
   \big[ \bar q'_{s} P_L  {b}_{v} \big]
  \big[ \bar d_{\bn,\omega_2}  \nslash P_L\,\cP^\beta_\perp q_{\bn,\omega_3} 
  \big]
  \big[ \bar q_{n,\omega_1}  \bnslash \gamma^\perp_\beta P_R\, q'_{n,\omega_4} 
   \big]
   \,, \nn\\
  O_{5d-8d}^{(2L)} &=  O_{1d-4d}^{(2L)}\,\, \frac{3e_{q'}}{2} \,,
\end{align}
with sums over $q,q'=u,d,s$. Note that the flavor structure of these operators
is identical to $O_{4d}^{(1L)}$. For the the electroweak penguin operators
$O_{7,8}$ an additional four operators $O_{5d-8d}^{(2L)}$ are needed, which have
the same spin-flavor structures as $O_{1d-4d}^{(2L)}$, but with an $e_{q'}$
charge factor, $\sum_{q,q'} 3e_{q'}/2$.  Again we caution that we have not
considered the complete set of local $\Lambda^2/m_b^2$ operators, since our
basis does not include three-body terms with an $ig {\cal B}_\perp^\mu$, nor
terms with an extra $D_s$ soft covariant derivative. We have also not considered
${\cal O}(\mu_{M_1}\mu_{M_2}\Lambda/m_b^3)$ terms. All these terms are real, and
it would be interesting to calculate them in the future.

The weak Hamiltonian with Wilson coefficients for the operators $O_{id}^{(2L)}$
is
\begin{equation} \label{match2}
 H_W =  \frac{4G_F}{\sqrt2}\, (\lambda^{(d)}_u + \lambda^{(d)}_c) \sum_{n,\bn} 
   \int [d\omega_{1}d\omega_{2}d\omega_{3}  d\omega_{4}]\, 
    \sum_{i=1-8} a_i^{\chi}(\omega_j)\,  O_{id}^{(2L)}(\omega_j) \,.
\end{equation}
Since only the penguin operators $O_{5-8}$ contribute, we pulled out the
common CKM factor.   Matching at tree level onto the operators $O_{id}^{(2L)}$ by
keeping terms linear in the $\perp$-momenta in Fig.~\ref{fig:Ann}, we find
\begin{align}\label{chimatch}
a^\chi_{1}(x,y) &= \frac{4 C_F \pi \alpha_s(\mu_h) }{N_c}\,  \bigg[ 
  \Big( C_6 + \frac{C_5}{N_c} \Big) F_1(x,y)
  + \frac{C_5}{N_c}\, 
  F_2(x,y)
 \bigg]_{\mbox{\o}}\,,
   \nonumber \\
a^\chi_{2}(x,y) &= \frac{4 C_F \pi \alpha_s(\mu_h) }{N_c}\,  \bigg[ 
  -\Big( C_6 + \frac{C_5}{N_c} \Big) 
  F_1(\bar y,\bar x)
  + \frac{C_5}{N_c}\, 
  F_2(\bar y,\bar x)
  \bigg]_{\mbox{\o}}\,,
  \nonumber \\
a^\chi_{3}(x,y) &= \frac{4 C_F \pi \alpha_s(\mu_h) }{N_c}\, \bigg[ 
  -\Big( C_6 + \frac{C_5}{N_c} \Big)
  F_3(x,y)
  - \frac{C_5}{N_c}\, 
 F_2(x,y)
 \bigg]_{\mbox{\o}}\,,
   \nonumber \\
a^\chi_{4}(x,y) &= \frac{4 C_F \pi \alpha_s(\mu_h) }{N_c}\, \bigg[ 
   \Big( C_6 + \frac{C_5}{N_c} \Big) 
  F_3(x,y)
  - \frac{C_5}{N_c}\, 
  F_2(\bar y,\bar x)
  \bigg]_{\mbox{\o}} \,,
 \nonumber\\[5pt]
a^\chi_{5-8}(x,y) &=   a^\chi_{1-4}(x,y) \quad
   \mbox{with } C_5\to C_7,\ C_6\to C_8 \,,
\end{align}
where $x$ and $y$ are defined in Fig.~\ref{fig:Ann} and
\begin{align}
  F_1(x,y) &= \bigg[\frac{1+\bar x}{y^2\, \bar y\, \bar x^2}\bigg]_{\mbox{\o}}
   \!\! + d_1(\mu_-) \delta'(\bar x) \bigg[ \frac{1}{y^2\bar y}\bigg]_{\mbox{\o}}
   \!\! + d_2(\mu_+) \delta'(y) \bigg[ \frac{1+\bar x}{\bar x^2}\bigg]_{\mbox{\o}}
   \!\! + d_3(\mu_\pm) \delta'(\bar x)\delta'(y) 
   \,,\nn\\
  F_2(x,y) &= \bigg[ \frac{1}{(1-x\bar y)\bar x y^2} \bigg]_{\mbox{\o}}
  \,,\nn\\
  F_3(x,y) &= \bigg[ \frac{1}{y^2\, \bar x^2}  \bigg]_{\mbox{\o}}
   \!\! + d_4(\mu_-) \delta'(\bar x) \bigg[ \frac{1}{y^2}\bigg]_{\mbox{\o}}
   \!\! + d_5(\mu_+) \delta'(y) \bigg[ \frac{1}{\bar x^2}\bigg]_{\mbox{\o}}
  \!\! + d_6(\mu_\pm) \delta'(\bar x) \delta'(y) 
  \,.
\end{align}  
Here $d_{1-6}$ play the same role as $d$ in Eq.~(\ref{F}).  The
coefficients $a^\chi_{1-8}$ are polluted in the sense of
Ref.~\cite{Bauer:2004tj}, meaning that ${\cal O}(\alpha_s^2)$ matching
results proportional to the large coefficients $C_{1,2}$ could compete
numerically. This makes the computation of these ${\cal
O}(\alpha_s^2)$ corrections important.

For decays involving a pseudoscalar in the final state, the operators
$O_{1d}^{(2L)}$ and $O_{2d}^{(2L)}$ generate so-called ``chirally enhanced''
terms, proportional to $\mu_M$.  Time-ordered products of \SCETa operators also
generate $\mu_M$ terms, but only at ${\cal O}(\alpha_s^2)$. It is not clear that
the chirally enhanced terms are larger numerically than other power corrections.
In particular three-body distributions from operators with $\bar\xi_n (ig {\cal
  B}_\perp^\mu) \Gamma \xi_n$ are parametrically (and sometimes numerically as
well) of similar importance~\cite{Chernyak:1983ej}. The distributions are
related by~\cite{Hardmeier:2003ig}
\begin{align} \label{3relation}
  f_P \mu_P \bigg[ \phi^{P\,\prime}_{\sigma}(x) 
   +\frac{(2x-1)}{x(1-x)}\, \phi_\sigma^P(x) \bigg]
   & = - {6 f_{3P}} \bigg[ \frac{G_{P_z}^{(t)}(x)}{x}
   +\frac{G_{P_y}^{(t)}(x)}{1-x} \bigg] \,,\nn\\
  f_P \mu_P \bigg[ \phi^{P}_{p}(x) - \frac{1}{6x(1- x)}\,
   \phi^P_\sigma(x)\bigg] 
   & = -{ f_{3P}} \bigg[ \frac{G_{P_z}^{(t)}(x)}{x}-\frac{G_{P_y}^{(t)}(x)}{1-x}
   \bigg] \,,
\end{align}
where $G_{P_z}^{(t)}(x)$ and $G_{P_y}^{(t)}(x)$ are integrals over the
three-parton distribution, $\phi_{3P}$. These relations allow certain chirally
enhanced terms with $\mu_P f_P$ to be traded for non-chirally enhanced terms
with $f_{3P}$. Thus it is clear that the chirally enhanced terms dominate over
the three-body operators only in the special case when the linear combinations
in the square brackets on the left-hand side of Eq.~(\ref{3relation}) are
numerically suppressed.  Solving with these linear combinations set to zero
determines the two-body distributions $\phi_\sigma^P$ and $\phi_p^P$ in the
Wandzura-Wilczek (WW) approximation~\cite{Wandzura:1977qf}. Thus in order to
uniquely specify the $\mu_P$ dependent terms, the WW approximation was needed in
Ref.~\cite{BBNS2}.

In contrast, in SCET we are not forced to assume a numerical dominance of the
$\mu_P$ terms to uniquely identify them.  We can instead define local chirally
enhanced annihilation terms to be the matrix elements of the operators
$O_{1d}^{(2L)}$ and $O_{2d}^{(2L)}$ for final states with a pseudoscalar. With a
minimal basis of operators, the matrix elements of these terms are unique. The
remaining terms involve other operators, and we postpone discussing them to
future work. We proceed to work out the factorization formula for
$O_{1d}^{(2L)}$ and $O_{2d}^{(2L)}$ with steps analogous to Eqs.~(\ref{factor})
through (\ref{Aann}).  To take the matrix element we need Eq.~(\ref{lcdistn})
and the result
\begin{equation} \label{phipp}
  \langle P_{n_1}(p) |\, \bar q^{(f)}_{n,\omega}\, \bnslash\,\,
  {\slash\!\!\!\! \cP_\perp} P_{R}\, q^{(f')}_{n,\omega'}\, | 0 \rangle 
  = - \frac{i}{6}\, c_{Pff'}\, \delta_{nn_1}\,
   \delta(\bn\mcdot p - \omega +\omega')\, f_P \mu_P\, \phi_{pp}^P(y) \,.
\end{equation}
Here $c_{Pff'}$ are Clebsch-Gordan factors, $y=\omega/\bn\cdot p$, and we have
not written the $\omega'$ dependence in the distribution due to the
$\delta$-function. The distribution $\phi^{P}_{pp}(y)$ is related to more
standard twist-3 two-parton and three-parton distributions
by~\cite{Hardmeier:2003ig,Manohar:2006nz}
\begin{equation} 
\phi^P_{pp}(y) = {3y} \bigg[
  \phi^P_p(y)+ \frac16\, \phi^{P\prime}_{\sigma}(y) + 
  \frac{2\, f_{3P}}{f_P \mu_P}\, \int \frac{dy'}{y'}\, 
  \phi_{3P}(y-y',y)\bigg] \,.
\end{equation}
Note that in $\mu_P \phi_{pp}^P$, the $\phi_{3P}$ term does not have the chiral
enhancement factor $\mu_P$. There will be additional terms proportional to
$\phi_{3P}$ generated by three-body operators.  We choose the $\phi_{pp}^P$ and
$\phi_{3P}$ basis of twist-three distributions, keeping in mind the relations in
Eq.~(\ref{3relation}).  For decays involving one or more pseudoscalars in the
final state we find the chirally enhanced local annihilation amplitudes
\begin{align} \label{Aann2}
A^{(2)}_{Lann} = 
  -\frac{G_F f_B  f_{M_1} f_{M_2}  }{6\sqrt{2}\, m_b}\, (\lambda^{(d)}_u +
  \lambda^{(d)}_c)\!
  & \int_0^1\!\!\! dx\, dy \Big[ \mu_{M_1}
   H_{\chi 1}(x,y)\, \phi_{pp}^{M_1}(y) \phi^{M_2}(x) 
  \nn\\
& + \mu_{M_2} H_{\chi 2}(x,y) 
    \phi^{M_1}(y) \phi^{M_2}_{pp}(x) \Big] ,
\end{align}
where $\mu_\rho=\mu_{K^*}=0$ and using isospin $\mu_\pi=m_\pi^2/(m_u+m_d)$,
$\mu_K=m_K^2/(m_s+m_u)=m_K^2/(m_s+m_d)$. Terms with $\phi_{3P}$ or terms of the
same order with a $D_s^\mu$ in their soft matrix elements have not been included
in our $A^{(2)}_{Lann}$, though they also give local annihilation contributions
to $A^{(2)}$. Furthermore, we focused on the pseudoscalar matrix element in
Eq.~(\ref{phipp}) to derive the contribution in Eq.~(\ref{Aann2}). The
$O_{1d,2d}^{(2L)}$ operators in Eq.~(\ref{Q2ann}) will contribute additional
terms for decays to longitudinal vector mesons involving distributions
$h_\parallel^{(s)'}$ and $h_\parallel^{(t)}$ (our notation for these
distributions follows Ref.~\cite{Hardmeier:2003ig}). The operators
$O_{3d,4d}^{(2L)}$ will produce decays to two transverse vectors with
distributions from among $\phi_\perp$, $F$, ${\cal V}$, ${\cal A}$. It would be
straightforward to work out a factorization theorem from the operators
$O_{id}^{(2L)}$ in terms of these distributions, though we will not do so here.

\begin{table}[t]
\begin{tabular}{|c|c|c|}
\hline\hline
$M_1 M_2$ & $H_{\chi 1}(x,y)$ & $H_{\chi 2}(x,y)$  
\\ \hline\hline 
  $\pi^0 \pi^-$, $\rho^0\pi^-$ $\pi^0 \rho^-$  &
   $-\frac{1}{\sqrt2}\, a^\chi_1(x,y)-\frac{1}{\sqrt2}\, a^\chi_5(x,y)$
   & $\frac{1}{\sqrt2}\, a^\chi_2(x,y)+\frac{1}{\sqrt2}\, a^\chi_6(x,y)$ \\
  $\pi^- \pi^0$, $\rho^-\pi^0$ $\pi^- \rho^0$
   & $\frac{1}{\sqrt2}\, a^\chi_1(x,y)+\frac{1}{\sqrt2}\, a^\chi_5(x,y)$ &
   $-\frac{1}{\sqrt2}\, a^\chi_2(x,y)-\frac{1}{\sqrt2}\, a^\chi_6(x,y)$ \\
$\pi^+ \pi^-$,  $\pi^+\rho^-$, $\rho^+\pi^-$ 
 &  $- a^\chi_1(x,y) +\frac{1}{2} a^\chi_5(x,y) $ 
  & $a^\chi_2(x,y) -\frac{1}{2} a^\chi_6(x,y)$ 
 \\
$\pi^0 \pi^0,$ $ \rho^0 \pi^0$ 
 & $  a^\chi_1(x,y)-\frac12\, a^\chi_5(x,y)  $ 
 &   $- a^\chi_2(x,y)+\frac12\, a^\chi_6(x,y) $ 
  \\
 $K^{-} K^{(*)+}$, $K^{(*)-} K^{+}$ &  
  {\bf ---} & {\bf ---}
  \\
  $\bar K^{0} K^{(*)0}$,  $\bar K^{(*)0} K^{0}$ 
  & $a^\chi_1(x,y)-\frac{1}{2} a^\chi_5(x,y)$ 
  &  $-a^\chi_2(x,y)+\frac{1}{2} a^\chi_6(x,y)$
  \\
 $K^{-} K^{(*)0}$, $K^{(*)-} K^{0}$ 
    & $a^\chi_1(x,y)+ a^\chi_5(x,y)$  
    &  $-a^\chi_2(x,y)- a^\chi_6(x,y)$
  \\
\hline
 $\pi^- \bar K^{(*)0}$, $\rho^- \bar{K}^{0}$ 
   & $  a_1^\chi(x,y)+a_5^\chi(x,y) $ & $-a_2^\chi(x,y)-a_6^\chi(x,y) $
  \\
 $\pi^0 K^{(*)-},\rho^0 K^-$  
   & $-\frac{1}{\sqrt 2}\, a_1^\chi(x,y)-\frac{1}{\sqrt 2}\, a_5^\chi(x,y) $ 
   & $\frac{1}{\sqrt 2}\, a^\chi_2(x,y) + \frac{1}{\sqrt 2}\, a^\chi_6(x,y) $ 
   \\
 $\pi^0 \bar K ^{(*)0},\rho^0 \bar K ^{0}$ 
   & $ \frac{1}{\sqrt 2} \, a_1^\chi(x,y)-\frac{1}{2\sqrt 2} \, a_5^\chi(x,y) $
   & $-\frac{1}{\sqrt 2} \, a^\chi_2(x,y)+\frac{1}{2\sqrt 2} \, a^\chi_6(x,y) $
   \\
 $\pi^+ K^{(*)-} ,\rho^+ K^{-}$ 
   &  $-a_1^\chi(x,y)+\frac{1}{2} a_5^\chi(x,y)$ 
   &  $a_2^\chi(x,y)-\frac{1}{2} a_6^\chi(x,y)$ \\
  \hline\hline
\end{tabular}
\caption{Hard functions for the annihilation amplitude $A_{Lann}^{(2)}$ in
Eq.~(\ref{Aann2}) for $\bar B^0$ and $B^-$ decays. The result for $B^- \to
\pi^0\pi^-$ is obtained by adding the results using the entries from
the first two rows, and so vanishes in the isospin limit.}
\label{table2a}
\end{table}
Results for the hard coefficients $H_{\chi 1}$ and $H_{\chi 2}$ in terms of the
Wilson coefficients $a_{i}^\chi$ are given in Table~\ref{table2a} for $\bar B^0$
and $B^-$ decays and in Table~\ref{table2b} for $\bar B_s$ decays. Note that
there are no chirally enhanced annihilation contributions for the $\bar B_s\to
\pi\pi$ or $\bar B_s \to \rho\pi$ channels, so $B_s$ decays could potentially be
used to separate annihilation contributions from $A^{(1)}_{Lann}$ and
$A^{(2)}_{Lann}$.  For later convenience we define moment parameters
\begin{align} \label{beta2}
  \beta_{\chi 1,\chi 5}^{M_1 M_2}  &= \frac{1}{6} \int_0^1\!\!\! dx\, dy \:
  a_{1,5}^{\chi}(x,y) 
  \, \phi_{pp}^{M_1}(y) \phi^{M_2}(x)   \,,
 \nn \\
   \beta_{\chi 2,\chi 6}^{M_1 M_2} &=  \frac{1}{6}  \int_0^1\!\!\! dx\, dy \:
  a_{2,6}^{\chi}(x,y)
   \phi^{M_1}(y) \phi^{M_2}_{pp}(x) \,.
\end{align}
Neglecting $\phi_{3P}$ in the WW approximation yields $\phi_{pp}^P(y)=6 y(1-y)$.
At order $\alpha_s(\mu_h)$ our results for $\beta_{\chi 1}$ and $\beta_{\chi
  2}$, taken with the WW approximation, agree with the convolutions derived in
this limit in Refs.~\cite{BBNS2,BN}. Ignoring the \o-distributions we would find
that these convolution integrals diverge.  The zero-bin avoided double counting
in our convolutions, and yields a finite and real result for the chirally
enhanced annihilation amplitude.

\begin{table}[t]
\begin{tabular}{|c|c|c|}
\hline\hline
$M_1 M_2$ & $H_{\chi 1}(x,y)$ & $H_{\chi 2}(x,y)$  
\\ \hline\hline 
$ K^+\pi^-$,  $K^{*+}\pi^- $, $K^+\rho^- $ 
  &  $- a^\chi_1(x,y)+\frac{1}{2} a^\chi_5(x,y)  $ 
  &  $a^\chi_2(x,y)-\frac{1}{2} a^\chi_6(x,y) $
 \\
$K^0\pi^0 $, $K^{*0}\pi^0 $, $K^0\rho^0 $
  & $\frac{1}{\sqrt 2}\, a^\chi_1(x,y) -\frac{1}{2\sqrt 2}\, a^\chi_5(x,y)  $  &  
  $-\frac{1}{\sqrt 2}\, a^\chi_2(x,y) +\frac{1}{2\sqrt 2}\, a^\chi_6(x,y)  $ 
  \\
\hline
 $K^{+}K^{-} $, $ K^{*+}K^{-}$, $K^{+}K^{*-} $
   &  $- a^\chi_1(x,y)+\frac{1}{2} a^\chi_5(x,y)$ & 
  $a^\chi_2(x,y) -\frac{1}{2} a^\chi_6(x,y)$ 
  \\
  $K^{0}\bar K^{0} $,  $K^{*0}\bar K^{0} $ , $ K^{0}\bar K^{*0}$ 
  &  $a^\chi_1(x,y)-\frac{1}{2} a^\chi_5(x,y)$ 
  & $-a^\chi_2(x,y)+\frac{1}{2} a^\chi_6(x,y)$ 
  \\
  \hline\hline
\end{tabular}
\caption{Hard functions for the annihilation amplitude $A_{Lann}^{(2)}$ in
Eq.~(\ref{Aann2}) for $\bar B_s$  decays.}
\label{table2b}
\end{table}

Let's see how the convolutions work out at order $\alpha_s(\mu_h)$ following
Ref.~\cite{Manohar:2006nz}.  We need two standard convolutions involving
zero-bin subtractions,
\begin{align}
\int_0^1\! dx\, dy\, \bigg[ \frac{1+\bar x}{y^2\, \bar y\,
  \bar x^2} \bigg]_{\mbox{\o}} \phi_{pp}^{M_1}(y) \phi^{M_2}(x) &= 
  \big\langle y^{-2}\, \bar y^{-1} \big\rangle^{M_1}_{pp}\, 
  \Big( \big\langle \bar x^{-2}\big\rangle^{M_2} 
   +\big\langle  \bar x^{-1} \big\rangle^{M_2} \Big)\,, \nn\\
\int_0^1\! dx\, dy\, \bigg[ \frac{1+y}{y^2\, x\, \bar x^2} \bigg]_{\mbox{\o}}
  \phi^{M_1}(y) \phi_{pp}^{M_2}(x) &= 
  \big\langle \bar x^{-2}\, x^{-1} \big\rangle^{M_2}_{pp}\,
  \Big( \big\langle y^{-2} \big\rangle^{M_1} 
   +\big\langle y^{-1} \big\rangle^{M_1} \Big) \,.
\end{align}
Here we model the $y^{-2}$, $y^{-1}$ moments as in Eq.~(\ref{Apif}) and
Eq.~(\ref{Apif2}), and for the remaining convolution we again assume there is no
interference between the rapidity renormalization and invariant mass
renormalization to find
\begin{equation} \label{Apif3}
\big\langle y^{-2}\,\bar y^{-1} \big\rangle^{M_1}_{pp} 
  = \int_0^1\!\! dy\, \bigg[\frac{\phi_{pp}^{M_1}(y,\mu)}{y^2(1\!-\!y)} - 
    \frac{y\phi_{pp}^{M_1\,\prime}(0,\mu)}{y^2} \bigg]
    + \phi^{M_1\prime}_{pp}(0,\mu)\, \ln\Big( \frac{n\mcdot p_{M_1}}{\muplus}\Big)
   \,.
\end{equation}
The $\mu_\pm$ dependence is canceled by tree level logarithmic dependence in the
coefficients, $d_{1,4}(\mu_-)=\ln(p_{M}^-/\mu_-)$,
$d_{2,5}(\mu_+)=\ln(p_M^+/\mu_+)$,
$d_{3,6}(\mu_\pm)=\ln(p_{M}^-/\mu_-) \ln(p_M^+/\mu_+)$.  The kernels
in Eq.~(\ref{chimatch}) also involve two more complicated convolutions
that are derived in Appendix~\ref{AppA},
\begin{align} \label{Apif4} 
& \big\langle [(1-x\bar y)\bar x y^2]^{-1}\big\rangle^{M_1M_2}_{pp} =
 \int_0^1\! dx\, dy \bigg[ \frac{1}{(1-x\bar y)\bar x y^2} 
   \bigg]_{\mbox{\o}}\, \phi_{pp}^{M_1}(y) \phi^{M_2}(x) \nn\\
& \quad  = \int_0^1\! dx  \int_0^1\! dy \bigg[
   \frac{ \phi_{pp}^{M_1}(y) \phi^{M_2}(x) }{(\bar x + y - \bar x y)\bar x y^2}
  - \frac{ \phi_{pp}^{M_1\,\prime}(0)\,\phi^{M_2}(x)}{(\bar x +  y)\bar x y}
  \bigg] 
  - \phi_{pp}^{M_1\,\prime}(0) \int_0^1 dx\,
  \frac{ \phi^{M_2}(x)\, \ln(2 - x)}{(1-x)^2} \,, \nn\\
&  \big\langle [(1-x\bar y)\bar x^2 y]^{-1}\big\rangle^{M_2M_1}_{pp} =
  \int_0^1\! dx\,dy \bigg[ \frac{1}{(1-x\bar y)\bar x^2 y} \bigg]_{\mbox{\o}}\,
  \phi^{M_1}(y) \phi_{pp}^{M_2}(x) \\
& \quad = \int_0^1\! dy \int_0^1\! dx \bigg[
  \frac{  \phi^{M_1}(y)\, \phi_{pp}^{M_2}(x)}{(\bar x + y - \bar x y)\bar x^2 y}
  + \frac{ \phi^{M_1}(y)\, \phi_{pp}^{M_2\,\prime}(1)}{(\bar x +  y)\bar x y} \bigg] 
  + \phi_{pp}^{M_2\,\prime}(1) \int_0^1 dy\,
   \frac{ \phi^{M_1}(y)\, \ln(1 + y)}{y^2} \,. \nn
\end{align}
As promised, the minimal subtraction scheme yields a well defined result for
$A_{Lann}^{(2)}$. The scheme dependence cancels order by order in $\alpha_s$
between the matrix element and perturbative corrections to the kernels obtained
by matching. In any scheme the result at order $\alpha_s(\mu_h)$ is real.


\section{Generating Strong Phases }
\label{sec:phase}


In this section we derive results for the order at which strong phases occur in
the power suppressed amplitudes $A^{(1)}$.  It is convenient to classify complex
contributions to the $B\to M_1 M_2$ amplitudes according to the distance scale
at which they are generated. We use the terminology hard, jet, and
nonperturbative to refer to imaginary contributions from the scales $m_b$,
$\sqrt{m_b\Lambda}$, and $\Lambda^2$ respectively.  We will not attempt to
classify strong phases generated by charm loops, since a complete understanding
of factorization for these terms order by order in a power counting expansion is
not yet available.

For a matrix element to have a physical complex phase it must contain
information about both final state mesons.  Generically, terms in the factorized
power expansion of $B\to M_1 M_2$ amplitudes involve only vacuum to meson matrix
elements, so strong phase information can be contained in the Wilson
coefficients or the factorized operators, but not in the states. This provides
tight constraints on the source of strong phases. Nonperturbative strong phases
will occur if matrix elements of these factorized operators give complex
distribution functions.  A sufficient condition to generate a nonperturbative
phase, is to have a factorized operator that is sensitive to the directions of
two or more final state mesons~\cite{Mantry:2003uz}, information that can be
carried by Wilson lines.  Physically, this is a manifestation of soft
rescattering of final states.  In processes like ours where soft-collinear and
collinear$(n)$-collinear$(\bar n)$ factorization are relevant, and there is only one hadron in any given light cone direction, this criterion
implies that all strong phases reside in the soft matrix elements, where the
directional information from collinear hadrons is retained in soft Wilson lines,
$S_r$, with direction $r^\mu$. Since $S_r^\dagger S_r =1$ these Wilson lines
often cancel, but for many of the power suppressed terms listed in
Table~\ref{table0} the cancellation is not complete.  This mechanism for
generating a strong phase was first observed for $\bar B^0 \to
D^0\pi^0$~\cite{Mantry:2003uz}, where a nonperturbative soft matrix element
occurs through four-quark operators depending on $n$ and $v'$ (which are null
and time-like vectors for the final state light and charmed mesons, respectively).

For the $B\to M_1 M_2$ decays with two energetic light mesons, a nonperturbative
strong phase requires a soft matrix element depending on the $S_n$ and $S_\bn$
Wilson lines in \SCETb.  The simplest way to obtain the Wilson lines for the
soft operators is to match \SCETa onto \SCETb~\cite{Bauer:2002aj}.  In \SCETa
one first uses the decoupling field redefinition on collinear
fields~\cite{SCET}, $\xi_n\to Y_n\xi_n$, $\xi_\bn \to Y_\bn \xi_\bn$, $A_n\to
Y_n A_n Y^\dagger_n$ and $A_\bn\to Y_\bn A_\bn Y^\dagger_\bn$, which generates
the Wilson lines and factorizes usoft and collinear fields.  The fields of a
given type are then grouped together by Fierz rearrangements.  Matching the
resulting operators or time-ordered products onto \SCETb gives $Y_r \to S_r$,
and we can read off which soft Wilson lines are present. Because of the
properties of the subleading \SCETa operators, we will not have an $S_n$ and
$S_\bn$ in the final \SCETb operator unless we have a subleading \SCETa
Lagrangian with an $n$-collinear field and usoft fields, and one with
$\bn$-collinear fields and usoft fields. We used this property to determine
which entries are real or complex, and listed the results in the last column of
Table~\ref{table0}. The complex entries with multiple ${\cal L}_{\xi
  q}^{(j)}$'s~\cite{bcdf} also have at least two hard-collinear gluons, and
so generate contributions that start at $\alpha_s(\mu_i)^2$ when matched onto
\SCETb.

To determine the perturbative order of the complex contributions, we must also
classify which hard and jet coefficients give complex phases. In general any
hard coefficient generated by matching at $\ge 1$ loop will give imaginary
contributions, since these loops involve fields for both final state mesons, as
pointed out for the general case in Ref.~\cite{BBNS} and for charm loops in
Ref.~\cite{Bander:1979px}.  Since all leading order contributions in
Table~\ref{table0} have at least one $\alpha_s(\mu_i)$, the hard imaginary
contributions for $A^{(0)}$ are ${\cal O}[\alpha_s(\mu_i)\alpha_s(\mu_h)/\pi]$. 
At order $\Lambda/m_b$ all annihilation contributions but $Q_i^{(4)}$ have at
least one $\alpha_s(\mu_i)$, and for these terms the hard complex contributions
involve $\alpha_s(\mu_i)\alpha_s(\mu_h)$ and thus are smaller than the
nonperturbative terms proportional to $\alpha_s(\mu_i)^2$. For $Q_i^{(4)}$ the
amplitude is real at the leading perturbative order, $\alpha_s(\mu_h)$, as
demonstrated in section~\ref{sec:local}, and so hard complex contributions start
at $\alpha_s^2(\mu_h)$. In contrast for the amplitude $A_{rest}^{(1)}$ a complex
amplitude is generated at order $\alpha_s(\mu_i)\, \Lambda/m_b$, which is only
suppressed by $\Lambda/m_b$ compared to $A^{(0)}$.

Finally, we should examine complex contributions from the jet scale. At leading
order there is a unique jet function $J$~\cite{Bauer:2004tj}. $J$ also
contributes to the heavy-to-light form factors and only knows about the
$n$-collinear direction. Thus $A^{(0)}$ does not get imaginary contributions at
any order in the $\alpha_s(\mu_i)$ expansion (which has been demonstrated
explicitly to $\alpha_s^2(\mu_i)$~\cite{Becher:2004kk}). At next-to-leading order
in the power expansion, there is no known relation of the power suppressed jet
functions with analogous jet functions in the form factors. However, the
subleading jet functions also depend only on one collinear direction, and do not
carry information about both final state mesons that could generate a physical
strong phase. We demonstrate this fact more explicitly by examining the
calculation at ${\cal O}(\alpha_s(\mu_i))$, which is sufficient to see that the
amplitudes are real up to the order where a nonperturbative phase first occurs.
At this order the jet functions are generated by matching tree level \SCETa
diagrams onto \SCETb.  A typical example is
\begin{equation}
  \frac{1}{(x+i\epsilon)\, (k^+ + i\epsilon)} \,,
\end{equation}
where $x$ is a momentum fraction that will be convolved with a collinear
distribution function, and the $k^+$ will be convolved with a soft distribution
function. These jet functions are real if and only if we can drop the
$i\epsilon$ factors.  However, just as in section~\ref{sec:local}, the
$i\epsilon$ terms can be dropped because the zero-bin
subtractions~\cite{Manohar:2006nz} ensure that this does not change the
convolution.\footnote{A equivalent physical argument for dropping the
  $i\epsilon$ factors was given in Ref.~\cite{Mantry:2003uz}, where it was
  needed to prove that certain long-distance contributions are absent in color
  suppressed decays.} Thus factorization gives real ${\cal
  O}(\alpha_s(\mu_i))$ jet functions.

This demonstrates that complex contributions in the power suppressed
annihilation amplitudes are suppressed,
\begin{equation}
  {\rm Im} \bigg[ \frac{A^{(1)}_{ann}}{A^{(0)}} \bigg] = {\cal O}\bigg(
  \frac{\alpha_s(\mu_i)}{\pi}\, \frac{\Lambda}{m_b}\bigg) 
  + {\cal O}\bigg( \frac{\Lambda^2}{m_b^2}
  \bigg) \,.
\end{equation}
On general grounds one might have expected ${\cal O}(\Lambda/m_b)$ suppressed
strong phases, which we have demonstrated are absent in $A_{ann}^{(1)}$, though
they do occur in $A^{(1)}_{rest}$.

\begin{figure}
  \includegraphics[width=5.5cm]{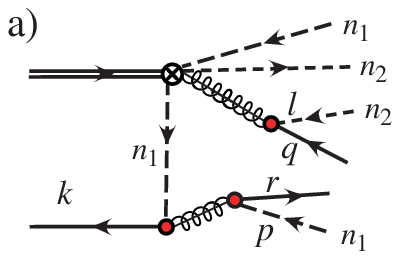} \qquad\quad
   \includegraphics[width=6cm]{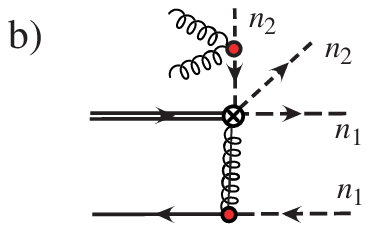} 
\caption{Graphs which generate a strong phase in lowest order matching of \SCETa
  operators onto \SCETb: a) has a $Q^{(1)}$, two ${\cal L}_{\xi_{n_1}
  q}^{(1)}$, and one ${\cal L}_{\xi_{n_2} q}^{(1)}$ and contributes to the
  annihilation amplitude at ${\cal O}(\alpha_s^2(\mu_i))$; and b) has a
  $Q^{(1)}$, one ${\cal L}_{\xi_{n_1} q}^{(1)}$, and one ${\cal L}_{\xi_{n_2}
  \xi_{n_2}}^{(2)}$ and contributes to non-annihilation amplitudes at ${\cal
  O}(\alpha_s(\mu_i))$. Dashed quark lines are $n_1$ or $n_2$ collinear, and
  solid quark lines are soft.}
\label{topann}
\end{figure}

We close this section by giving two examples of time-ordered products generating
the nonperturbative strong phases discussed above. We consider a time-ordered
product with three ${\cal L}^{(1)}_{\xi q}$ insertions contributing to
annihilation. When matching onto \SCETb we integrate out the hard-collinear
modes, leading to an eight-quark operator. Figure~\ref{topann}a shows the order
$\alpha_s^2(\mu_i)$ contribution to this matching. The soft quark lines remain
open as their contraction leads to an on-shell line which must be treated
nonpertrubatively.  The resulting \SCETb operator has the generic form
 \begin{align}
O^{II} &= J(n_2\cdot p,n_1 \cdot l,n_1\cdot r,n_2\cdot q,n_1 \cdot k) \\
&\quad \times ( \bar q_s S_{n_1})_{n_1\cdot r}\, \Gamma^{(1)}
  (S_{n_2}^\dagger q_s)_{n_2\cdot q}\, (\bar q_s S_{n_2})_{n_1 \cdot k}\, 
  \Gamma^{(2)} (S^\dagger_{n_1} h_v)\,
  (\bar q_{n_1,l} \Gamma^{(3)} q_{n_1,l'})\,
  (\bar q_{n_2,p'}\Gamma^{(4)} q_{n_2,p}) \nn
\end{align}
where we use the shorthand subscript notation, $(S_{n_i}^\dagger q_s)_{n_i\cdot
  q} \equiv [\delta(n_i\mcdot q-n_i\mcdot \cP) S_{n_i}^\dagger q_s]$.  We took
the jet directions to be $n_1$ and $n_2$, rather than $n$ and $\bar n$, to
emphasize that the soft operator is sensitive to the relative directions of the
jets.  The functions $S_i$ shown in Table~\ref{table0} are defined by the matrix
element of this type of operator
\begin{equation} \label{Si}
S_i(n_1 \mcdot k,n_1\mcdot r,n_2\mcdot q,)\equiv \langle 0 |
  (\bar q_s S_{n_1})_{n_1\cdot r}\, \Gamma_i^{(1)} 
  (S_{n_2}^\dagger q_s)_{n_2\cdot q}\, (\bar q_s S_{n_2})_{n_1 \cdot k}\,
  \Gamma_i^{(2)}  (S^\dagger_{n_1} h_v) | B(v) \rangle\,,
\end{equation}
where $i$ runs over color, Dirac, and flavor structures.  To count the factors
of $\pi$ in these amplitudes, note that the hard-collinear contractions give
$g^4$, and that the matrix element of the resulting four-quark operator,
$\langle 0 | (\bar q \ldots q)( \bar q \ldots b_v) | B\rangle$, is suppressed by
$1/(4\pi)^2$ relative to $\langle 0 | (\bar q \ldots b_v) | B\rangle$. (The
four-quark operator has an extra loop with no extra couplings.) This
demonstrates that nonperturbative complex contributions first occur at order
$[\alpha_s(\mu_i)^2/\pi] (\Lambda/m_b)$, i.e., suppressed by
$[\alpha_s(\mu_i)/\pi] (\Lambda/m_b)$ compared to the leading amplitudes. The
phases arising from the type of matrix element shown in Eq.~(\ref{Si}) play a
crucial role in explaining the observed strong phases which arise in color
suppressed decays~\cite{Mantry:2003uz}. Their resulting operators predict the
equality of amplitudes and strong phases between decays involving $D$ and $D^*$
mesons and have been confirmed in the data~\cite{Pirjol:2004yf}.  This type of
diagrams also have long-distance contributions of the same order, which arise
from time-ordered products in \SCETb and can also be complex.  To see this note
that the hard-collinear quark propagator in Fig.~\ref{topann}a could also be
on-shell (i.e., have ${\cal O}(\Lambda^2)$ virtuality), in which case it would
remain open until the matrix element is taken at the low scale.  By opening that
line we see that this contribution corresponds to the time-ordered product of a
four-quark operator and a six-quark operator, both of which are generated when
matching onto \SCETb. A long-distance part is the same order in
$\alpha_s(\mu_i)$ and does not change our conclusions about these terms.  In
Fig.~\ref{topann}b we show a non-annihilation contribution to $\hat
A_{rest}^{(1)}$ which is of order $\alpha_s(\mu_i)\Lambda/m_b$. This term is
generated by the time-ordered product of $Q^{(1)}$, an insertion of the
$n_1$-collinear ${\cal L}_{\xi q}^{(1)}$, and an operator with $n_2$-collinear
quarks and usoft gluons,
\begin{equation}
{\cal L}_{\xi\xi}^{(2)}= (\bar\xi_n W)\,  Y_{n_2}^\dagger i
  \Dslash_{us}^\perp\, i\Dslash_{us}^\perp Y_{n_2}\,
  \frac{\bnslash}{2\bnP}\, (W^\dagger \xi_n) \,.
\end{equation}


\section{Applications and Conclusion}
\label{sec:phen}


\subsection{Phenomenological Implications}

To understand the implications of the experimental data, it is crucial to know
which contributions to the $B\to M_1M_2$ amplitudes can be complex.  The best
sensitivity to non-SM physics is via interference phenomena, where new
interactions enter linearly (instead of quadratically), such as $CP$-violating
observables.  The sensitivity to such effects depends on how well we understand
the dominant and subdominant SM amplitudes, including their strong phases.  The
existence of strong phases in $B$ decays is experimentally well established
(e.g., the $B\to D\pi$ and $B\to \pi\pi$ rates, the $CP$ asymmetry
$A_{K^+\pi^-}$, the transversity analysis in $B\to J/\psi K^*$, etc.).

One example of how strong phase information can be useful is the method for
determining $\gamma$ from $B\to \pi\pi$ proposed in Ref.~\cite{Bauer:2004dg}.
The method uses isospin, the factorization prediction that ${\rm Im}(C/T) \sim
{\cal O}\big(\alpha_s(m_b), {\Lambda}/{m_b} \big)$, and does not require data on
the poorly measured direct $CP$ asymmetry $C_{\pi^0\pi^0}$.\footnote{Here $C$
  and $T$ are isospin amplitudes defined in the $t$-convention, where
  $\lambda_t$ is eliminated from the amplitudes in favor of $\lambda_c$ and
  $\lambda_u$.}  The phases in $A^{(0)}$ at $\alpha_s(m_b)\alpha_s(\mu_i)$ are
calculable and partially known~\cite{BBNS,Beneke:2005vv}.  The current $B\to
\pi\pi$ data is in mild conflict (at the $\sim2\sigma$ level) with the SM CKM
fit~\cite{Grossman:2005jb}.  More precise measurements are needed to understand
how well the theoretical expectations are satisfied, and to decipher whether
there might be a hint for new physics. Obviously further information about power
corrections in ${\rm Im}(C/T)$ could help to clarify the situation. 

In all factorization-based approaches to charmless $B$ decays, several
parameters are fit from the data or are allowed to vary in certain ranges. The
choice and ranges of these parameters should be determined by the power
counting. This motivated keeping the charm penguin amplitudes, $A_{c\bar c}$ as
free parameters in SCET~\cite{Bauer:2004tj}, as was done earlier in
Ref.~\cite{Ciuchini}.  In the BBNS approach these are argued to be
factorizable~\cite{BBNS}. A fit to the data using this parameterization found
large power suppressed effects~\cite{Charles:2004jd} including annihilation
amplitudes, which might be interpreted as a breakdown of the $\Lambda/m_b$
expansion. In QCD sum rules, the annihilation amplitude was found to be of the
expected magnitude and to have a sizable strong
phase~\cite{Khodjamirian:2005wn}, but a distinction between the terms we
identify as real local annihilation and complex time-ordered product
annihilation was not made.

Channels like $B\to K\pi$ and $B\to K\bar K$ are sensitive to new physics, but
by the same token are dominated by penguin amplitudes, which can have charm
penguin, annihilation, and other standard model contributions. Since there are
possible large nonperturbative $c$-loop contributions in $A_{c\bar c}$ that have
the same $SU(3)$ flavor transformation properties as annihilation terms, they
cannot be easily distinguished by simple fits to the data.  However, in a
systematic analysis based on SCET these correspond to different operators'
matrix elements, so it is possible to disentangle the various contributions and
determine their expected size. The factorization theorems for annihilation
amplitudes derived here only involve distributions that already occurred at
leading order. This means that we can compare the size of annihilation
amplitudes to experimental data without further ambiguities from additional
hadronic parameters. We take up this comparison in section~\ref{sec:model}
below.

As an explicit example of how to assemble our results in
sections~\ref{sec:local} and~\ref{sec:chiral}, we derive the local annihilation
amplitude for $\bar B^0\to K^-\pi^+$. From Table~\ref{table1a} we can read off
the result for this channel, $H(x,y)=-a_4^s(x,y)-a_8^s(x,y)$, and from
Table~\ref{table2a}, $H_{\chi 1}=-a_1^\chi(x,y)+1/2\, a_5^\chi(x,y)$ and $H_{\chi
  2}=a_2^\chi(x,y)-1/2\, a_6^\chi(x,y)$. With the lowest order matching results in
Eqs.~(\ref{aLO}) and (\ref{chimatch}) we can set $a_8=0$ and $a_{4u}=a_{4c}$,
which inserted into Eqs.~(\ref{Aann}) and (\ref{Aann2}) gives
\begin{align}
 A_{Lann}^{(1)}(K^-\pi^+) &= \frac{G_F f_B f_{\pi} f_{K} }{\sqrt{2}}\,
 (\lambda_c^{(s)}\!+\!\lambda_u^{(s)}) \int\!\! dx \, dy\, a_{4u}(x,y) \phi^\pi(y)
 \phi^K(x) \\
  &= \frac{G_F f_B f_{\pi} f_{K} }{\sqrt{2}}\,
 (\lambda_c^{(s)}\!+\!\lambda_u^{(s)})\, \beta_{4u}^{\pi K}\,, \nn\\
 A_{Lann}^{(2)}(K^-\pi^+) &= \frac{G_F f_B f_{\pi} f_{K} }{6\sqrt{2}}\,
 (\lambda_c^{(s)}\!+\!\lambda_u^{(s)})  \int\!\! dx \, dy\, \Big[
 \frac{\mu_\pi}{m_b}\, \Big\{a_1^\chi(x,y)-\frac12 a_5^\chi(x,y)\Big\}
   \phi_{pp}^\pi(y) \phi^K(x) \nn\\
 &\qquad\qquad
  - \frac{\mu_K}{m_b}\, \Big\{ a_2^\chi(x,y) -\frac12 a_6^\chi(x,y)\Big\} \phi^\pi(y) \phi_{pp}^K(x) \Big] \nn\\
 &= \frac{G_F f_B f_{\pi} f_{K} }{\sqrt{2}}\,
 (\lambda_c^{(s)}\!+\!\lambda_u^{(s)})  \Big[
 \frac{\mu_\pi}{m_b}\, \Big\{  \beta_{\chi 1}^{\pi K}
  -\frac{1}{2} \beta_{\chi 5}^{\pi K} \Big\}
  - \frac{\mu_K}{m_b}\, \Big\{ \beta_{\chi 2}^{\pi K} 
   -\frac12 \beta_{\chi 6}^{\pi K} \Big\} \Big] \,. \nonumber
\end{align}
Thus, both the leading order annihilation amplitude $A^{(1)}_{Lann}$,
and the chirally enhanced annihilation amplitude $A^{(2)}_{Lann}$ are
determined by the $\beta$'s defined in Eqs.~(\ref{beta1}) and
(\ref{beta2}). Other $K\pi$ channels have similar expressions with
different Clebsch-Gordan coefficients. To the local annihilation
contributions we must add the hard-collinear annihilation terms
computed in Ref.~\cite{Arnesen2}, $A_{\rm hard-collin}^{(1ann)}$,
since they are the same order in $\alpha_s$ and $1/m_b$ as the
$A_{Lann}^{(1)}$ terms. To see explicitly what the $\beta$'s involve
we insert the ${\cal O}(\alpha_s)$ values of $a_{3u}(x,y)$,
$a_1^\chi(x,y)$, and $a_2^\chi(x,y)$ to give
\begin{align} \label{Kpi}
  & A_{Lann}(K^-\pi^+) =  -\frac{G_F f_B f_{M_1} f_{M_2} }{\sqrt{2}} 
  (\lambda_c^{(s)}+\lambda_u^{(s)}) \frac{4\pi\alpha_s(\mu_h)}{9}  
  \nn\\
 &\quad \times \bigg\{ \Big(\frac{C_9}{6}-\frac{C_3}{3}\Big) \Big[
  \big\langle \bar x^{-2}\big\rangle^K \big\langle y^{-1} \big\rangle^\pi 
 - \big\langle [y(x\bar y-1)]^{-1} \big\rangle^{\pi K}
  + d(\mu_-) \phi_K'(1)\langle y^{-1} \big\rangle^\pi  \Big] \nn\\
 & \qquad
 -\frac{2\mu_\pi}{3m_b} 
  \Big({C_6}\!-\!\frac{C_8}{2}\!+\!\frac{C_5}{3}\!-\!\frac{C_7}{6}\Big) \Big[ 
  \big\langle y^{-2} \bar y^{-1} \big\rangle_{pp}^{\pi} 
  \big(\big\langle \bar x^{-2}\big\rangle^K \!+\! \big\langle \bar
   x^{-1}\big\rangle^K \big)
   + d_1(\mu_-) \phi_K'(1)  \big\langle y^{-2} \bar y^{-1}
   \big\rangle_{pp}^{\pi} \nn\\ 
 & \qquad \ \ \
   - d_2(\mu_+) \phi_\pi'(0) \big(\big\langle \bar x^{-2}\big\rangle^K \!+\! \big\langle \bar
   x^{-1}\big\rangle^K \big) - d_3(\mu_\pm)  \phi_K'(1) \phi_\pi'(0) \Big]
 \nn\\
 &\qquad
  - \frac{2\mu_\pi}{3m_b} 
\Big(\frac{C_5}{3}\!-\!\frac{C_7}{6}\Big)
   \big\langle [(1-x \bar y) \bar x
   y^2]^{-1}\big\rangle^{\pi K}_{pp}
 +\frac{2\mu_K}{3m_b} \Big(\frac{C_5}{3}\!-\!\frac{C_7}{6}\Big)
   \big\langle [(1-x\bar y) \bar x^2 y]^{-1}
  \big\rangle_{pp}^{K\pi} 
 \nn\\
& \qquad
  -\frac{2\mu_K}{3m_b} 
   \Big({C_6}\!-\!\frac{C_8}{2}\!+\!\frac{C_5}{3}\!-\!\frac{C_7}{6}\Big)
  \Big[ \big(\big\langle y^{-2} \big\rangle^\pi+ \big\langle y^{-1}
  \big\rangle^\pi \big) 
 \big\langle x^{-1}\bar x^{-2} \big\rangle^K_{pp} 
   - d_1(\mu_+) \phi_\pi'(0)  \big\langle \bar x^{-2} x^{-1}
   \big\rangle_{pp}^{K} \nn\\ 
 & \qquad \ \ \
   + d_2(\mu_-) \phi_K'(1) \big(\big\langle y^{-2}\big\rangle^\pi \!+\! \big\langle
   y^{-1}\big\rangle^\pi \big) - d_3(\mu_\pm)  \phi_\pi'(0) \phi_K'(1) \Big]
  \bigg\} .
\end{align}
Here results for the convolutions denoted by brackets $\langle \cdots
\rangle$ can be found in Eqs.~(\ref{Apif}), (\ref{Apif2}),
(\ref{Apif3}), and (\ref{Apif4}) in the minimal subtraction scheme.
Results for other channels can be assembled in a similar fashion.
Corrections to $A_{Lann}+A_{\rm hard-collin}^{(1ann)}$ are suppressed
by ${\cal O}[\alpha_s^2(\mu_i) /(\pi\alpha_s(m_b))]$, while we caution
that additional $\alpha_s(\mu_h)\Lambda/m_b$ terms without a $\mu_\pi$
or $\mu_K$ will be present in the last two lines of
Eq.~(\ref{Kpi}). In the next subsection we derive results for all of
these channels using a simple model for the distribution functions,
and study numerically the size of the annihilation amplitudes.

Annihilation contributions have been claimed to play important roles in several
observables~\cite{Keum,Keum:2000wi,BBNS2,BN,alex}, in particular in generating
large strong phases in $B\to K\pi$ decays~\cite{Keum,Keum:2000wi}.  The $B\to
\pi\pi$ and $K\pi$ data indicate that the latter decays are dominated by penguin
amplitudes, and the pattern of rates and $CP$ asymmetries is not in good
agreement with some predictions.  In particular, it is not easy in the BBNS
analysis to accommodate the measured $CP$ asymmetry, $A_{K^+\pi^-} = -0.108 \pm
0.017$~\cite{HFAG}, except in the S3 and S4 models of Ref.~\cite{BN}.  In these
models the annihilation contributions are included by using asymptotic
distributions, and divergent integrals are parameterized as $\int_0^1 dx/x \to
X_A$ and $\int_0^1 dx \ln x/x \to -X_A^2/2$, with $X_A = (1 + \varrho_A
e^{i\varphi_A}) \ln(m_B/500\, {\rm MeV})$. Model S3 postulates $\varrho_A = 1$,
$\varphi_A = -45^\circ$ for all final states, while in the S4 scenario
$\varrho_A = 1$ and $\varphi_A = -55^\circ,\, -20^\circ,\, -70^\circ$ for the
$PP,\, PV,\, VP$ channels, respectively. Thus
\begin{equation}
S3: \ X_A = 4.0 -1.7\,i \,, \qquad
   S4: \ X_A = \{ 3.7-1.9\,i\:,\: 4.6-0.8\,i \:,\: 3.2-2.2\,i \} \,.
\end{equation} 
In addition, $\alpha_s(\mu)$ and the Wilson coefficients are evaluated at the
$\mu_i$ intermediate scale~\cite{BN}.  

Our result for the factorization of annihilation contributions derived in
Sec.~\ref{sec:local} constrains models of annihilation.  Equation~(\ref{Aann})
gives a well defined and real amplitude at leading order, which depends on
twist-2 distributions, $\phi_M$. It does not involve model parameters
$\varrho_A$ and $\varphi_A$.  For $A^{(1)}_{Lann}$ using Eq.~(\ref{Apif}) and
the asymptotic form of the meson distributions, we find a correspondence
\begin{equation}
``X_A" = 1 + \int_0^1\! dx\, \frac{\phi_\pi(x)}{6\, (x^2)_{\mbox{\o}}}
  = \ln\Big( \frac{m_b}{\muplus}\Big) \,.
\end{equation}
Clearly, $X_A$ is real. The asymptotic distributions $\sim 6x(1-x)$ are more
accurate for large scales, and at the matching scale where $\mu_+\sim m_b$,
$X_A$ is not enhanced by a large logarithm. This matching scale $\mu_+$ should
not be decreased below $m_b$ since $\mu_+\sim m_b$ is already the correct scale
for collinear modes with $p^+\sim m_b$. We estimate $|X_A| \lesssim 1$.
Thus, the modeling of annihilation contributions with complex $X_A$ in the BBNS
approach (including the phenomenologically favored S3 and S4 scenarios)
are in conflict with the heavy quark limit, and should be constrained to give
smaller real $X_A$'s.

In the KLS~\cite{Keum} treatment of annihilation, complex amplitudes are
generated from dynamics at the intermediate scale from the $i\epsilon$ in
propagators. The MS-factorization used in the derivation of our annihilation
amplitudes demonstrates that including the $i\epsilon$ term in collinear
factorization would induce a double counting. Thus we expect such contributions
to physical strong phases to be realized by operators with soft exchange that
occur at higher order in $\Lambda/m_b$ and therefore to be small.

Annihilation contributions were also argued to play an important role in
explaining the large transverse polarization fraction in $B\to\phi
K^*$~\cite{alex}.  It was shown that factorization implies $R_T = {\cal
  O}(1/m_b^2)$, where $R_T$ denotes the transverse polarization
fraction~\cite{alex}.  Subsequently, it was shown using SCET that $R_T$ is power
suppressed unless a long-distance charm penguin amplitude $A_{c\bar c}$ spoils
this result~\cite{Bauer:2004tj,Williamson:2006hb}.  Experimentally, one finds
$R_T(B\to \phi K^*) \approx 0.5$~\cite{HFAG}, while $R_T(B\to \rho\rho)$ is at
the few percent level. It has been argued that the large $R_T(B\to \phi K^*)$
may provide a hint of new physics in the $b\to s\bar s s$ channel.  In
Ref.~\cite{alex} it was suggested that standard model annihilation contributions
may account for the observed large value of $R_T(B\to \phi K^*)$. Our analysis
in Sec.~\ref{sec:chiral} agrees with~\cite{alex} in that annihilation
contributions to the transverse polarization amplitude at first order in
$\alpha_s$ are suppressed by not one, but two powers of $\Lambda/m_b$. However,
we do not find a numerical enhancement of these terms (which in~\cite{alex} is
partly due to the large sensitivity of the $(2X_A-3)(1-X_A)$ function to
$\varrho_A$ in the BBNS parameterization). The operators in Eq.~(\ref{Q2ann})
give rise to transverse polarization, but since MS-factorization renders the
naively divergent convolutions finite, these power suppressed amplitudes do not
receive sizable enhancements.  Although we have not derived explicit results for
the $B\to \phi K^*$ annihilation amplitudes (since $\phi$ is an isosinglet), our
results make it unlikely that local annihilation can explain the $R_T(B\to \phi
K^*)$ data. We have not explored whether the time-ordered products at ${\cal
  O}(\alpha_s^2(\mu_i) \Lambda/m_b)$ could give rise to transverse polarization,
and it would be interesting to do so.

\subsection{\boldmath Annihilation amplitudes with simple models for
$\phi^M(x)$ and $\phi_{pp}^M(x)$}  \label{sec:model}

In this section we derive numerical results for the local annihilation
amplitudes in various channels using a simple model for the distributions. It is
convenient to write the $\Delta S=0$ local annihilation amplitude as
\begin{align} \label{Lann}
  A_{Lann}(\bar B\to M_1 M_2) &= -\frac{G_F f_B f_{M_1} f_{M_2} }{\sqrt{2}} \:
  \bigg\{ \lambda^{(d)}_u  h_u(\bar B\to M_1M_2) 
   + \lambda^{(d)}_c  h_c(\bar B\to M_1M_2) \nn \\
 &\hspace{-1cm}
   +  (\lambda^{(d)}_u + \lambda^{(d)}_c) \Big[ \frac{ \mu_{M_1} }{m_b} \:
   h_{\chi 1}(\bar B\to M_1 M_2)
   + \: \frac{ \mu_{M_2} }{m_b} \: h_{\chi 2}(\bar B\to M_1M_2) 
   \Big] \bigg\} \,. 
\end{align} 
For $\Delta S=1$ decays we replace $\lambda^{(d)}_{u,c}\to \lambda^{(s)}_{u,c}$.
The coefficients $h_u$, $h_c$, $h_{\chi 1}$, and $h_{\chi 2}$ are equal to
linear combinations of $\beta_{iu}$, $\beta_{ic}$, $\beta_{\chi 1}$,
$\beta_{\chi 2}$, $\beta_{\chi 5}$, and $\beta_{\chi 6}$ with Clebsch-Gordan
coefficients determined from Tables~\ref{table1a}, \ref{table1b}, \ref{table2a},
\ref{table2b}. The combinations are simply determined by the replacements
\begin{align}
 h_u &= \big(H(x,y) \mbox{ with } \tilde a_i^{d,s}(x,y) \to \beta_{iu}^{M_1M_2},\
   \tilde a_i^{d,s}(y,x) \to \beta_{iu}^{M_2M_1} \big) 
  \,,
   \nn\\
  h_c &= \big(H(x,y) \mbox{ with } \tilde a_i^{d,s}(x,y) \to \beta_{ic}^{M_1M_2},\
   \tilde a_i^{d,s}(y,x) \to \beta_{ic}^{M_2M_1} \big)
  \,,
  \nn\\
  h_{\chi 1} &= \big(H_{\chi 1}(x,y) \mbox{ with }  a_{1,5}^{\chi}(x,y) \to \beta_{\chi
    1,\chi 5} \big)
  \,,\nn\\
   h_{\chi 2} &= \big(H_{\chi 2}(x,y) \mbox{ with }  a_{2,6}^{\chi}(x,y) \to \beta_{\chi
     2,\chi 6} \big)
  \,.
\end{align}
For the coefficients $a_{3u,3c,4u,4c,7,8}$ and the $a_i^\chi$'s, the ${\cal
  O}(\alpha_s^2 C_{1,2})$ matching corrections could be comparable numerically
with the ${\cal O}(\alpha_s C_{3-10})$ corrections considered here. This should
be kept in mind when examining numbers quoted below for the corresponding
$\beta$'s.

Results for the coefficients $\beta_{iu}$, $\beta_{ic}$, and $\beta_{\chi i}$,
can be found in Eqs.~(\ref{beta1}) and (\ref{beta2}).  To derive numerical
results we need to model the meson distribution functions.  We take the $C_i$
from Eq.~(\ref{Ci}), use
\begin{align}
  \alpha_s(\mu_h) &= 0.22\,, 
 & \mu_\pi(\mu_h) &=  2.3\,{\rm GeV}\,,
 & \mu_K(\mu_h) &= 2.7\,{\rm GeV}\,,\nn\\
 f_K &= 0.16\,{\rm GeV}, 
 & f_\pi &=0.13\,{\rm GeV}, 
 & f_B  &= 0.22\,{\rm GeV},
\end{align}
where $\mu_h=m_b=4.7\,{\rm GeV}$, $f_B$ comes from a recent lattice
determination~\cite{Gray:2005ad}.  For the $\phi$'s we take simple models with
parameters $a_i^M$ and $a_{ipp}^M$ which we consider specified at
the high scale $\mu_h$,
\begin{align}
  \phi^M(x) &= 6x(1-x) \big[1+ a_1^M (6x-3) + 6 a_2^M ( 1- 5 x+ 5x^2)\big]
  \,,\nn\\
  \phi_{pp}^P(x) &= 6x(1-x) \big[1+ a_{1pp}^P (6x-3) + 6 a_{2pp}^P ( 1- 5
  x+ 5x^2)\big] \,.
\end{align}
Based on recent lattice data for moments of the $\pi$ and $K$
distributions~\cite{Braun:2006dg} we take $a_2^{\pi,K}=0.2\pm 0.2$, where the
lattice error was doubled to give some estimate for higher moments.  For the
$\pi$ we set $a_1^\pi=a_{1pp}^\pi=0$, while for the $K$ we
use~\cite{Braun:2006dg} $a_1^K=0.05\pm 0.02$. We also take \(w_{3\pi,K}=-3\pm
1\), $a_{2pp}^{\pi,K}=0\pm 0.4$ and $a_{1pp}^K=0.0\pm 0.2$. Note that the range
for our parameters is similar to those used in the BBNS models~\cite{BBNS2,BN}
and light-cone sum rules~\cite{Ball:2006wn}. Since the uncertainties in the
model parameters are large and not significantly affected by variation of the
$\mu_\pm$ scales we keep these fixed at $m_b$, where the logs in the
$d_i(\mu_\pm)$ terms drop out and the constant under the logs are neglected. A
scan over models with parameters in these limits gives predictions for the
annihilation coefficients.  For the $\bar B\to K\pi$ channels we find
\begin{align} \label{betaKpi}
 \beta_{2u}^{\pi K} &= 1.8\pm 1.2,
 & \beta_{4u}^{\pi K} &=\beta_{4c}^{\pi K} =-0.15\pm 0.10, 
 & \beta_{2c}^{\pi K} &=0.14\pm 0.09
 \,,\nn\\
  \beta_{hc 1}^{\pi K} &=0.09\pm 0.33,
 & \beta_{hc 2}^{\pi K} &= -0.29\pm 0.09,
 \ \ \beta_{hc 3}^{\pi K} =-0.012\pm 0.002,
 & \beta_{hc 4}^{\pi K} &= 0.002\pm 0.01
\,,\nn\\
  \beta_{\chi 1}^{\pi K} &=0.0\pm 6.5 ,
 & \beta_{\chi 2}^{\pi K} &= 0.0\pm 5.8,
 \ \ \beta_{\chi 5}^{\pi K} =0.0\pm 0.094 ,
 & \beta_{\chi 6}^{\pi K} &= 0.0\pm 0.11
  \,.
\end{align}
Using these numbers we can compare the size of the local annihilation amplitudes to
the $\bar B\to K^-\pi^+$ data,
\begin{align} \label{RA1}
   R_A(K^-\pi^+) &=\frac{|A^{(1)}_{Lann}(K^-\pi^+)+A^{(2)}_{Lann}(K^-\pi^+)|}
    {|A_{Expt. Penguin}(K\pi)|} = 0.11\pm  0.09 \,,
  \nn\\ 
   R_A(\bar K^0\pi^-) &=
   \frac{|A^{(1)}_{Lann}(\bar K^0\pi^-)+A^{(2)}_{Lann}(\bar K^0\pi^-)|}
    {|A_{Expt. Penguin}(K\pi)|} = 0.12\pm  0.09 \,.
\end{align}
For the numerator we did a Gaussian scan using the values from
Eq.~(\ref{betaKpi}), and determined the error by the standard
deviation.  For the denominator we used the experimental penguin
amplitude determined by a fit to the $B\to K\pi$ data in
Ref.~\cite{Bauer:2005kd}.  Numerical results for annihilation
amplitudes with three-body distribution functions were considered in
Ref.~\cite{Arnesen2}. Although they are similar in size to
$A^{(1)}_{Lann}$ they cause only a $\sim 10\%$ change in the value of
$R_A(K^-\pi^+)$ in Eq.~(\ref{RA1}). The values of $R_A$ indicate that a fairly
small portion of the measured penguin amplitude is from
annihilation. We do not quote values for the ratio
$A_{Lann}^{(2)}/A_{Lann}^{(1)}$, since each of the numerator and
denominator can vanish and the parametric uncertainties are very
large. For typical values of the parameters in the $K\pi$ channels we
find that the $A_{Lann}^{(2)}$ is comparable or even larger than
$A_{Lann}^{(1)}$ in agreement with Ref.~\cite{BBNS2}.  The size of the
annihilation amplitudes in Eq.~(\ref{RA1}) are consistent with our
expectation for these power corrections.  For $B\to \bar K K$ we find
\begin{align}
  \beta_{1u}^{\bar K K} & =-9.6\pm 6.2, 
 & \beta_{2u}^{\bar K K} & =1.7\pm 1.1,
 & \beta_{3u}^{\bar K K}&=\beta_{3c}^{\bar K K} =0.63\pm 0.37,
 \nn\\
  \beta_{4u}^{\bar K K}&=\beta_{4c}^{\bar K K} =-0.14\pm 0.09, 
 & \beta_{1c}^{\bar K K} &= -0.03\pm 0.02, 
 & \beta_{2c}^{\bar K K} & = 0.13\pm 0.08,
 \nn\\
  \beta_{3u}^{K\bar K} &=\beta_{3c}^{K\bar K} =0.63\pm 0.37 ,
 & \beta_{\chi 1}^{\bar K K} &=0.0 \pm 6.5 ,
 & \beta_{\chi 2}^{\bar K K} &= 0.0\pm 5.5 
  \nn\\
  \beta_{\chi 5}^{\bar K K} &=0.0 \pm 0.095 ,
 & \beta_{\chi 6}^{\bar K K} &= 0.0\pm 0.11 
  \,.
\end{align}
Using these results to determine the $\lambda_c^{(d)}$ annihilation
contributions to $B\to \bar K K$ and comparing this to the experimental penguin
amplitude from Ref.~\cite{Bauer:2005kd} gives
\begin{align}
   R_A(K^-K^0) &=\frac{|A^{(1)}_{Lann}(K^-K^0)+A^{(2)}_{Lann}(K^-K^0)|}
    {|A_{Expt. Penguin}(\bar K K)|} =  0.15\pm 0.11\,.
%
\end{align}
This is similar in size to the ratios $R_A(K^-\pi^+)$, $R_A(\bar K^0\pi^-)$ and
so also consistent with a power correction.

\subsection{Conclusions}

In summary, we exhibited how a new factorization in SCET renders the
annihilation and ``chirally enhanced" annihilation contributions finite in
charmless nonleptonic $B\to M_1 M_2$ decays to non-isosinglet mesons.  We
constructed a complete basis of \SCETb operators for local annihilation
contributions as well as factorization theorems valid to all orders in
$\alpha_s$. By matching the full QCD diagrams onto \SCETb operators we showed
that their matrix elements are real at leading order in $\Lambda/m_b$ and
$\alpha_s(m_b)$.  The lowest order annihilation contributions depend on $f_B$
and a modified type of twist-2 distributions $\phi^{M_{1,2}}$ with dependence on
rapidity cutoffs.  Chirally enhanced local annihilation contributions depend in
addition on modified distributions $\phi^{M_{1,2}}_{pp}$. The annihilation
contributions can only have an unsuppressed complex part at ${\cal
  O}(\Lambda/m_b)$ if perturbation theory at the intermediate scale,
$\sqrt{\Lambda m_b}$, breaks down.

In the previous literature models for the power suppressed annihilation
corrections were often found to give enhanced contributions with large strong
phases, and such assumptions have been important in some fits to the data.
Considering all power suppressed amplitudes not involving charm loops, we proved
that complex annihilation contributions only occur suppressed by
$\alpha_s(\sqrt{\Lambda m_b})\, \Lambda_{\rm QCD}/m_b$ compared to the leading
amplitudes. From our factorization theorem we found that annihilation
contributes $(11\pm 9)$\% of the penguin amplitude in $\bar B^0\to K^-\pi^+$,
$(12\pm 9)$\% in $B^-\to \bar K^0\pi^-$, and $(15\pm 11)$\% in $B^-\to K^-K^0$.
We anticipate that our results will guide future fits to the vast amount of data
on charmless $B$ decays, and yield a better understanding of what this data
means.

\acknowledgments

This work was supported in part by the Director, Office of Science, Offices of
High Energy and Nuclear Physics of the U.S.\ Department of Energy under the
Contract DE-AC02-05CH11231 (Z.L.), the cooperative research agreement
DOE-FC02-94ER40818 (C.A. and I.S.), and DOE-ER-40682-143 and DEAC02-6CH03000
(I.R.).  I.S.~was also supported in part by the DOE OJI program and by the Sloan
Foundation.

\appendix

\section{Zero-bin subtractions for a two-dimensional distribution}
\label{AppA}

In this appendix we derive a result for the action of the zero-bin subtractions
on the integrand obtained from the chirally enhanced annihilation computation,
shown in Eq.~(\ref{Apif4}). Since the result involves a correlation in the
$x$ and $y$ integrals it cannot be read off from the results in
Ref.~\cite{Manohar:2006nz}. It is convenient to write the momentum fraction
factor coming from the offshell $b$-quark propagator as $(1-x\bar y)=(\bar x + y
-\bar x y)$. Including the rapidity convergence factors~\cite{Manohar:2006nz},
the integral we need is
\begin{equation} \label{A1}
I = \sum_{x\ne 1,\, y\ne 0} \int\! dx_r dy_r\, 
   \frac{ \phi_{pp}^{M_1}(y) \phi^{M_2}(x)}{(\bar x+ y- \bar x y)\bar x y^2}\,
   \Theta_x\Theta_y\, | x (1-x)|^\epsilon\, | y (1-y)|^\epsilon\,
   \Big(\frac{\mu_+\mu_-}{\bn\mcdot p_{1}\, n\mcdot p_{2} }\Big)^{2\epsilon} \,,
\end{equation}
where $\Theta_x=\theta(x)\theta(1-x)$. To determine the subtraction terms we
must look at the singular behavior as we scale towards the $x=1$ and $y=0$ bins,
which we do by taking $\bar x\sim\eta$ and $y\sim\eta$. In this limit the gluon
and $b$-quark in Fig.~\ref{fig:Ann} become soft, and this region would be double
counted without the zero-bin conditions. First consider the denominator,
\begin{equation}
 \frac{1}{\bar x + y -\bar x y}= \frac{1}{(\bar x +y)} + \frac{\bar x y}{(\bar
   x+y)^2} + \ldots \,.
\end{equation}
In the first term the $x$ and $y$ dependence does not decouple, so we must
consider them simultaneously. All terms beyond the first one produce finite
integrals and are dropped in the minimal subtraction scheme. For the numerator
in Eq.~(\ref{A1}) we use $\phi_{pp}^{M_1}(0)=\phi^{M_2}(1)=0$ and expand
\begin{align}
  \phi_{pp}(y)\, \phi(x) 
 &= -y \phi_{pp}^{\prime}(0)\, \bar x  \phi^\prime(1) 
   - \frac{y^2}{2} \phi_{pp}^{\prime\prime}(0)\, \bar x \phi^\prime(1) 
   + y \phi_{pp}^{\prime}(0)\, \frac{\bar x^2}{2}\, \phi^{\prime\prime}(1) +
   \ldots \nn\\
 &=  y \phi_{pp}^{\prime}(0) \sum_{n=1}^\infty \frac{(-\bar x)^n}{n!}\, \phi^{(n)}(1) 
  - \frac{y^2}{2}\, \phi_{pp}^{\prime\prime}(0)\, \bar x \phi^\prime(1) +\ldots \,.
\end{align}
In the first term on the last line we have identified all terms which remain
singular when multiplied by $1/[\bar x y^2 (\bar x\!+\! y)]$. This term is equal
to $y \phi_{pp}^\prime(0) \phi(x)$. Taken together with the expansion of
$\Theta_x\Theta_y$ we therefore find that the required minimal subtraction is
\begin{equation}
  \frac{y \phi_{pp}^{M_1\prime}(0) \phi^{M_2}(x)}{(\bar x+ y)\bar x y^2}\,
   \Theta_x\, \theta(y) \,.
\end{equation}
Following Ref.~\cite{Manohar:2006nz} we use this to convert Eq.~(\ref{A1}) into
an integral that includes the $x=1$ and $y=0$ regions,
\begin{align} 
I &= \int_0^1\!\! dx \int_0^1 \!\! dy \bigg[
   \frac{ \phi_{pp}^{M_1}(y) \phi^{M_2}(x)}{(\bar x+ y-\bar x y)\bar x y^2}
  - \frac{y \phi_{pp}^{M_1\prime}(0) \phi^{M_2}(x)}{(\bar x+ y)\bar x y^2}
  \bigg] \nn\\
  & \qquad - \int_0^1\!\! dx \int_1^\infty \!\! dy\,  
   \frac{y \phi_{pp}^{M_1\prime}(0) \phi^{M_2}(x)}{(\bar x+ y)\bar x y^2}\,
   x^\epsilon (1-x)^\epsilon\, y^\epsilon (y-1)^\epsilon\,
   \Big(\frac{\mu_+\mu_-}{\bn\mcdot p_{1}\, n\mcdot p_{2} }\Big)^{2\epsilon} \\
&= \int_0^1\!\! dx\, \frac{\phi^{M_2}(x)}{\bar x} \int_0^1\!\! dy \bigg[
   \frac{ \phi_{pp}^{M_1}(y) }{(\bar x+ y-\bar x y) y^2}
   - \frac{ \phi_{pp}^{M_1\prime}(0)}{(\bar x+ y) y} \bigg]
  - \int_0^1\!\! dx\! \int_1^\infty\!\!\! dy\, 
   \frac{\phi_{pp}^{M_1\prime}(0) \phi^{M_2}(x)}{(\bar x+ y)\bar x y}\,
   y^\epsilon (y-1)^\epsilon \nn \\
&= \int_0^1\!\! dx\, \frac{\phi^{M_2}(x)}{\bar x} \int_0^1\!\! dy \bigg[
   \frac{ \phi_{pp}^{M_1}(y) }{(\bar x+ y-\bar x y) y^2}
   - \frac{ \phi_{pp}^{M_1\prime}(0)}{(\bar x+ y) y} \bigg]
  - \phi_{pp}^{M_1\prime}(0) \int_0^1\!\! dx\,
   \frac{ \phi^{M_2}(x) \ln(2-x)}{(1-x)^2} \,. \nn 
\end{align}
Here in simplifying the term carrying the $y\to\infty$ limit, we noted that 
the integral is finite, and so it does not induce $\mu_\pm$ dependence in our
subtraction scheme. This result for $I$ was used in Eq.~(\ref{Apif4}).  For the
asymptotic pion wave functions, $\phi^\pi(x) = 6x(1-x)$ and $\phi_{pp}^\pi(y) =
6y(1-y)$, we obtain
$I = 36 + 6\pi^2 - 144 \ln 2\, = -4.60$. Note that the steps used here to derive the
subtraction also give the correct result for cases where the $x$ and $y$
integrals factorize, such as an integrand $\phi(x)\phi(y)/(x^2 y^2)$.

\end{document}